\documentclass[two column]{aa}  
\usepackage{ulem}
\usepackage{amsmath}

\newcommand{\unit}[1]{\,\rm{#1}}
\newcommand{\lqq}{\!<\!}
\newcommand{\rqq}{\!>\!}

\usepackage{graphicx}
\usepackage{soul,censor}
\newlength\nextcharwidth
\makeatletter
\renewcommand\@cenword[1]{%
  \setlength{\nextcharwidth}{\widthof{#1}}%
  \censorrule{\nextcharwidth}%
  \kern -\nextcharwidth%
  #1}
\makeatother

\usepackage{txfonts}
\usepackage[colorlinks = true,
linkcolor= blue,
citecolor = blue,
filecolor = blue,
urlcolor = blue,
]{hyperref}

\usepackage{booktabs,dcolumn}
\newcolumntype{P}[1]{>{\centering\arraybackslash}p{#1}}

\begin{document}

\title{Astrometric mass measurement of compact companions in binary systems with Gaia}
\author{
Yilun~Wang\inst{1,2}
\and Shilong~Liao\inst{3,2}
\and Nicola~Giacobbo\inst{4}
\and Aleksandra~Olejak\inst{5}
\and Jian~Gao\inst{6}\fnmsep\thanks{Corresponding author: Jian~Gao}
\and Jifeng~Liu\inst{2}\fnmsep\thanks{Corresponding author: Jifeng~Liu}
}

\institute{National Astronomical Observatories, Chinese Academy of Sciences, Beijing 100101, China\\
\email{wyl2013@mail.ustc.edu.cn}
\and School of Astronomy and Space Sciences, University of Chinese Academy of Sciences, Beijing 100049, China\\
\email{jfliu@nao.cas.cn}
\and Shanghai Astronomical Observatory, Chinese Academy of Sciences, 80 Nandan Road, Shanghai 200030, P.R.China
\and School of Physics and Astronomy \& Institute for Gravitational Wave Astronomy, University of Birmingham, Birmingham, B15 2TT, United Kingdom
\and Nicolaus Copernicus Astronomical Center, Polish Academy of Sciences, ul. Bartycka 18, 00-716 Warsaw, Poland
\and Department of Astronomy, Beijing Normal University, Beijing 100871,China\\
\email{jiangao@bnu.edu.cn}
}
\date{Received 31 March 2022. Accepted 27 June 2022.}
\abstract
   {For binary systems with an unseen primary and a luminous secondary, the astrometric wobble of the secondary could be used to study the primary. With Gaia, it is possible to measure the mass of the black hole or neutron star   with a luminous companion (hereafter BH/NS-LC).} 
   {Our aim is to provide a method for predicting  Gaia's ability in  measuring the mass  of  BH/NS-LCs. We also tried to estimate the number of solvable BH/NS-LCs using Gaia.}
   {We used a realistic Markov chain Monte Carlo simulation of mock Gaia observations to obtain a relation between the uncertainty of mass measurement of the primary in BH/NS-LCs with the observable variables of the secondary astrometric orbit. Furthermore, we used the \textsf{MOBSE} code to evolve a Galactic BH/NS-LC sample with a combined Milky Way model. Our relation is applied to this sample to estimate the number of solvable BH/NS-LCs.}
   {We derived a good relation between the mass uncertainty and the binary parameters. For the first time, we show the quantitive influence of the period $P$, inclination $i$, eccentricity $e$, and  ecliptic latitude $\beta$ to the mass measurement. Our results suggest that $48^{+7}_{-7}$ BH-LCs and $102^{+11}_{-10}$ NS-LCs are solvable during a $5\unit{yr}$ Gaia mission. We also give the distribution of the distance and apparent magnitude of the Gaia solvable BH/NS-LCs. This solvable sample would be increased by additional spectroscopic data or a prolonged Gaia mission.}
  {The mass uncertainty relation could be used in future simulations of BH/NS-LCs observed by Gaia. The prediction of the solvable BH/NS-LCs is not only influenced by the process in generating the Galactic BH/NS-LC sample, but is also affected by our uncertainty relation. In particular, the relations of parameters such as $[P,e,i,\beta]$ are very useful to correct the selection effect in the statistic results of the future BH/NS-LC sample observed by Gaia.}
\keywords{
astrometry ---
methods:numerical ---
stars:binaries:general ---
stars:black holes
}
\titlerunning{}
\authorrunning{}
\maketitle
\section{Introduction}
The first space astrometric mission, Hipparcos, observed about 120 thousand sources with an effective distance up to $1\,\rm{kpc}$ in 1989 \citep{hipparcos}. Of these sources,  18 thousand  are recorded as non-single stars, 235 of which come with a binary orbit solution \citep{hip_annex}. After more than 20 years, its successor Gaia has already enlarged the astrometry sample at least 12 thousand times, providing over 1.46 billion sources with full astrometric data and 344 million sources with only mean positions in the Gaia Early Data Release 3 \citep{astrometry_solution_edr3}. 

There are two steps in measuring astrometric binaries, detecting the potential astrometric binary systems and then solving their orbits. In the first step the basic astrometric information of current releases of Gaia data are obtained by treating all of them as single stars \citep{core_solution}. It is still possible to search for astrometric binaries from such astrometric solution by using  Renormalised Unit Weight Error ($RUWE$) and $excess\_noise$ \citep{belokurov2,belokurov1,keivan2021}. Furthermore, using the proper motion anomaly, it is also possible to characterise the presence of  companions with stellar or substellar mass \citep{pma_binary}. 

In the second step there have been various efforts to solve the orbital parameters of the binary systems using the Gaia astrometric epoch data. \cite{pourbaix_prediction} suggests that Gaia could obtain the orbit solution for 8.8 million unresolved binaries. Gaia is also able to obtain the binary orbit solution for BHs, NSs, or white dwarfs with a luminous companion (BH/NS/WD-LCs)  \citep{wp_gaia_stellar}, or even to grab the signal of exoplanets in some nearby sources \citep{gaia_planet}. \cite{Andrews2019} (hereafter A19) made a realistic Markov chain Monte Carlo (MCMC) simulation of such BH/NS/WD-LC systems with mock Gaia epoch data. Their result shows that the orbit of BH-LCs can be solved up to about $1\unit{kpc}$, while the uncertainty of NS-LCs can reach $0.1\,{M_{\odot}}$ at this distance.

To predict the number of Gaia solvable BH/NS-LC systems, people need not only the Gaia observation constraints, but also the model of binary evolution and Milky Way stellar population synthesis. Binary evolution codes like Massive Objects in Binary Stellar Evolution (\textsf{MOBSE}), \textsf{StarTrack}, and Stellar EVolution N-body (\textsf{SEVN}) could provide the evolution track of a binary system from birth to death \citep{kick1,kick2,startrack,sevn}. For Milky Way stellar populations, dozens of  papers are listed in \cite{greg1}, \cite{Constrain_2M0521}, and \cite{olejak2020}(hereafter O20). Combining the Gaia observation constraints with the simulated  Milky Way BH-LC population, \cite{revealing} suggests that 3,800-12,000 BH-LCs could be discovered by Gaia at the end of the  five-year mission. \cite{greg1} proposed that the number of Gaia solvable BH-LCs is in the range  41 to 340, after further studying Gaia's observation capability. Other works not only give the prediction of dozens to hundreds of Gaia detectable BH/NS-LCs, but also compare the results between different evolution models \citep{Gould2002,Yalinewich2018,yamaguchi2018,revealing,chatterjee2021,shikauchi2020,shikauchi2021,Constrain_2M0521}. With the MCMC  simulation of Gaia observations, A19 applied their constraints to a simulated BH/NS-Giant population, and estimated that there are  74 BH-Gs and 190 NS-Gs with a relative precision better than 0.3, which are somewhat solvable in Gaia's vision. This relative precision is defined by Equation 14 in A19, which is parametrised by the angular size of the orbital separation and the number of Gaia observations for each source in a five-year mission.

In this paper,  to study whether a BH/NS-LC source is solvable with Gaia observation, we focus on the second step. We used the  MCMC simulation method to explore Gaia's ability in a wider parameter range and studied the Gaia-solvable BH/NS-LCs with a simulated Galactic BH/NS-LC population. In Sect.~\ref{section:model} we try to construct a realistic observation data model for astrometric binary systems. We give a useful expression for the relative error of the dark companion, which can be used to test Gaia's ability directly. In Sect.~\ref{section:MCMC_sim} we describe our simulation in detail, and provide a full error-prediction model. We apply it to our binary synthesis population in Sect.~\ref{section:Gaia_solvable_population}, which is constructed following the instructions in O20. In Section \ref{section:conclusion} we present our conclusions.

\section{Mock astrometric data}\label{section:model}

First, we introduce the scanning observation mode of Gaia in Sect.~\ref{section:observation_mode} and the kinematics of binary systems in Sect.~\ref{section:binary_kinematics}. Then, in Sect.~\ref{section:mass_measurement} we derive the theoretical relation between the uncertainty of mass measurement and the observable variables. Finally, we describe the method of generating the mock data in Sect.~\ref{section:mock_data}.

\subsection{Gaia observations}\label{section:observation_mode}

\begin{figure}[h]
\centering
\includegraphics[scale=0.67]{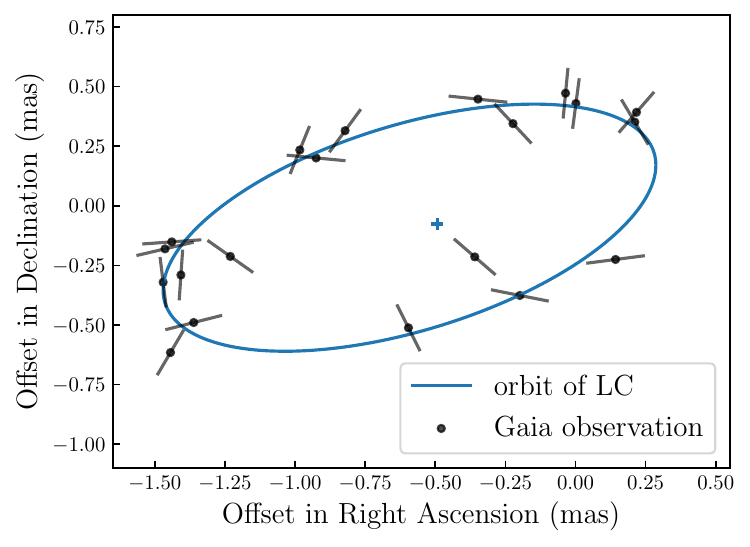}
\caption{Schematic diagram of the sky-projected orbit of the luminous component and the Gaia epoch data. The central blue cross is the barycentre. The blue line shows the binary orbit of the luminous star. The black points are the mock Gaia epoch data, which show offsets from the orbit. These offsets are caused by the uncertainty in AL direction, indicated by the orientation of the error bar. All the points omit the proper motion and the annual motion.}
\label{fig:toy_model}
\end{figure}

Gaia is a scanning satellite. When a star passes across these CCDs, the information of astrometry, photometry, and spectroscopy is  recorded \citep{gaia_mission}. During a $5\unit{yr}$ mission, each source could have an average of 75    astrometry observations. The satellite scans the sky at a fixed speed following its own scanning law, which makes the observation uneven and different at each place in the sky. There are two directions of the CCDs, the along-scan direction (AL) and the across-scan direction (AC)\footnote{For more details, please see Sect. 3.3.2 of \cite{gaia_mission}.}. In Fig.~\ref{fig:toy_model} we provide a schematic diagram of the mock Gaia observation of the luminous star in a fictitious NS-LC system\footnote{This figure is inspired by Figure 1 from the Gaia official website, https://www.cosmos.esa.int/web/gaia/iow\_20220131}. 

We followed the instructions in   Appendix A of \cite{perryman2014} that describes the stellar motion in the  AL direction\footnote{The motion along the AL direction is quite complicated;  we   referred to \cite{core_solution}, especially their Sects. 3.4 and 3.5.} rather than the right ascension-declination (RA-DEC) coordinates used in A19. The method of \cite{perryman2014} allows us to use Gaia observation auxiliary data directly. The displacement $\eta$ of a source in the AL direction on the tangent plane can be described by the following equations: 
\begin{eqnarray}\label{eq:complete_motion}
\eta(t)&=&[\Delta\alpha_0\cos\delta+\mu_{\alpha}\cos\delta{\cdot}(t-t_0)+B\cdot X+G\cdot Y]\sin\theta \nonumber\\
&&+[\Delta\delta_{0}+\mu_{\delta}{\cdot}(t-t_0)+A\cdot X+F\cdot Y]\cos\theta\nonumber\\ &&+\Pi_{\eta}\varpi.
\end{eqnarray}
Here $\theta$ is the position angle of the scanning direction and $\Pi_{\eta}$ is the along-scan parallax factor. The Gaia Observation forecast Tool (\textsf{GOST}) can give us the mock Gaia data of each point in the sky, containing $\theta$, $\Pi_{\eta}$, and the time $t$. In Eq.~\ref{eq:complete_motion} we used five parameters, $[\mathbf{\Delta\alpha_0\cos\delta},\Delta\delta_{0},\mathbf{\mu_{\alpha}\cos\delta},\mu_{\delta},\varpi]$, to describe the system motion. Except for the parallax $\varpi$, $[\mathbf{\Delta\alpha_0\cos\delta},\Delta\delta_{0}]$ stands for the start point along the RA-DEC direction, while $[\mathbf{\mu_{\alpha}\cos\delta},\mu_{\delta}]$ gives the proper motion. The orbital motion of the luminous component in a BH/NS-LC system are calculated by six intermediate parameters $[A,B,F,G,X,Y]$, which are introduced in Sect.~\ref{section:binary_kinematics}.

It is possible to obtain the radial velocity (RV) data from the archive or follow-up observations; these data are very helpful to restrict the binary orbital parameters. Gaia itself is also able to provide enough RV data since it can obtain the astrometric, photometric, and spectropic data at the same time. At the end of the 4-7 CCD rows on the  Gaia satellite is an area for collecting spectra \citep{gaia_mission}. Although a source $(V{<}15\unit{mag})$ is supposed to get 40 transits on this area during the $5\unit{yr}$ observation period, the actual number might be smaller, due to many  filters which decrease the number of spectra \citep{KatzD}. So we analysed the sources that have RV data in Gaia DR2. The average brightness of these sources is $V=12\,\rm{mag}$, whose precision could reach $1\,\rm{km\ s^{-1}}$. The average number of astrometry observations of these sources within 22 months is 30, which corresponds to 8.3 in spectroscopic data. This leads to 20 spectra for sources that have 75 astrometry observations. In this paper the position we used only has 62 astrometry observations in the $5\unit{yr}$ mission, thus it would only get 17 spectra. 

\subsection{Binary kinematics}\label{section:binary_kinematics}
We used Thiele-Innes elements $[A,B,F,G,X,Y]$ in Eq.~\ref{eq:complete_motion} to describe the luminous secondary motion in an ellipse orbit around the barycentre. In \cite{double_stars}, the Thiele-Innes elements are defined by the following equations:
\begin{eqnarray}\label{Thiele-Innes}
  A&=&a_s\cdot(\cos\omega\cdot \cos\Omega-\sin\omega\cdot \sin\Omega\cdot \cos i),\nonumber\\
  B&=&a_s\cdot (\cos\omega\cdot \sin\Omega+\sin\omega\cdot \cos\Omega\cdot \cos i),\nonumber\\
  F&=&a_s\cdot (-\sin\omega\cdot \cos\Omega-\cos\omega\cdot \sin\Omega\cdot \cos i),\\
  G&=&a_s\cdot (-\sin\omega\cdot \sin\Omega+\cos\omega\cdot \cos\Omega\cdot \cos i),\nonumber\\
  X&=&\cos E-e,\nonumber\\
  Y&=&\sqrt{1-e^2}\cdot \sin E.\nonumber
\end{eqnarray}
Here $a_s$ is the semi-major axis of the secondary expressed in radians, $e$ is the eccentricity, $i$ is the inclination, $\omega$ and $\Omega$ are used to describe the rotation between the true orbit and the projection orbit, $P$ is the period, and $T_p$ is the epoch of periastron passage. The parameter $E$ is the eccentric anomaly used to describe the position of the secondary on the orbit, which is defined by
\begin{equation}
   E=e\sin E+\frac{2\pi}{P}(t-T_p).\nonumber
\end{equation}

The $RV$ is mainly determined with $[K, e, P,\omega,\gamma]$, where $K$ and $\gamma$ are the Keplerian speed and the system radial velocity. The Keplerian speed $K$ of the secondary is
\begin{equation}
K=\frac{2\pi a_s \sin i}{P\varpi \sqrt{(1-e^2)}}.
\end{equation}
Then the $RV$ can be calculated as
\begin{equation}\label{eq:radial_velocity}
RV=\gamma+K[\cos(f+\omega)+e\cdot \cos\omega],
\end{equation}
where $f$ is the true anomaly:
\begin{eqnarray}
f=2\arctan[\sqrt{\frac{1+e}{1-e}}\tan(\frac{E}{2})].\nonumber
\end{eqnarray}

\subsection{Mass measurement}\label{section:mass_measurement}
If we solve the binary orbit from the mock data, then we can get the following relation for the total mass $M_{tot}$ from the third Keplerian law:
\begin{equation}
\label{eq:total_mass}
M_{tot}=m_p+m_s=\frac{4\pi^2(R_p+R_s)^3}{GP^2}.
\end{equation}
The subscripts $p$ and $s$ stand for primary and secondary, respectively, and $R$ is the length of the semi-major axis of the star orbit (in AU). Combining Eq.~\ref{eq:total_mass} with the relation between $R_s$ and $R_p+R_s$ yields
\begin{eqnarray}
\label{eq:Rs}
R_s=\frac{a_s}{\varpi}=\frac{m_p}{m_p+m_s}\cdot (R_p+R_s).\nonumber
\end{eqnarray}
We then define a new parameter $KF$:
\begin{equation}\label{eq:kepler_factor}
KF\equiv\frac{m_p^3}{M_{tot}^2}=\frac{4\pi^2}{G\varpi^3}\frac{a_s^3}{P^2} .
\end{equation}
In Eq.~\ref{eq:kepler_factor} the expression $m_p^3 M_{tot}^{-2}$ is proportional to $a_s^3P^{-2}$, which is similar to the right side of Eq.~\ref{eq:total_mass}, the third Keplerian law, hence we call it the  Keplerian factor $KF$. A19 studied the relative error of $KF$ and used it as a criterion to show whether a BH/NS-LC system is solvable. If the error is smaller than 0.3, they think the system is solvable. However, $KF$ contains both $M_{tot}$ and $m_p$, so its relative error is influenced strongly by the mass ratio $\frac{m_s}{m_p}$. Thus, using the error propagation method and Eq.~\ref{eq:kepler_factor}, we derive the relative error of $m_p$, 
\begin{eqnarray}\label{eq:mass_error_propagation}
\frac{\sigma(m_p)}{m_p}=
\frac{1}{m_p+3m_s}\cdot[(m_p+m_s)\cdot\Xi+2\sigma(m_s)],
\end{eqnarray}
which is a more direct variable and easier to interpret. In Eq.~\ref{eq:mass_error_propagation} we use the notation $\sigma(x)$ to present the error, which is half of the difference between the  16th and 84th percentiles, rather than the standard deviation $\sigma_x$. We also define a new variable $\Xi\,{\equiv}\,\frac{\sigma(KF)}{KF}$ for the relative error of $KF$ for simplicity in the rest of this article.

In the simulation $m_p$, $m_s$, and $\sigma(m_s)$ are fixed input parameters, while $\Xi$ is obtained from directly observable binary parameters. Equation \ref{eq:mass_error_propagation} helps us get the relative error of $m_p$.

\subsection{Generating the mock data}\label{section:mock_data}
To generate the mock data we needed to set the input parameters, generate the AL direction data using Eq.~\ref{eq:complete_motion}, and add the mock error to the data. We used Eq.~\ref{eq:radial_velocity} to generate the RV data.  

Some of the input parameters always stay the same for simplicity, for example $[\Delta\alpha_0\cos\delta,\Delta\delta_0,\omega,\Omega]$, which are fixed at $[0.1\unit{mas},0.2\unit{mas},11.5^{\circ},11.5^{\circ}]$. Similar to the values of $[0\unit{mas},0\unit{mas},30^{\circ},30^{\circ}]$ in A19, the values of these parameters are selected manually and kept the same for consistency in the whole study. Other parameters, such as $[P,a_s,\varpi,e,i,\beta]$, are varied in Sect.~\ref{section:MCMC_sim} to explore the parameter space. Here, $\beta$ is the barycentric ecliptic latitude. A19 shows that different proper motion values might have different influence on a binary system. So we analysed a Gaia DR2  subsample containing all sources that satisfy $8\lqq m_G\lqq14$, and obtained the mean proper motion when the parallax is $[10,1,0.5,0.25]\unit{mas}$. Here, $m_G$ is the $G$ band apparent magnitude of a source. At these parallax values we set the proper motion $\mu$ as $[56.8,7.1,4.9,4.7]\unit{mas{\cdot}yr^{-1}}$, respectively, rather than the uniform $10\unit{mas\cdot yr^{-1}}$ used in A19. The different components of proper motion $[\mu_{\alpha}\cos\delta,\mu_{\delta}]$ are $[\mu{\cdot}\sin{\frac{\pi}{6}},\mu{\cdot}\cos{\frac{\pi}{6}}]$.
In order to use Eq.~\ref{eq:complete_motion} to generate the AL direction data, we use \textsf{GOST} to get the observation information at $[\alpha_0,\delta_0]{=}$[6:11:49.07, 22:49:32.68]. It is the position of LB-1 \citep{LB1}, a wide star--black hole binary system candidate. Our initial purpose is simulating the astrometric motion of LB-1, and then extending it to  a full simulation.

We added the error $\rho_{AL}$ to $\eta$, following a normal distribution $\rho_{AL}{=}\mathcal{N}(0,\sigma_{AL})$, where $\sigma_{AL}$ is the standard deviation of $\eta$, mostly fixed at $0.1\unit{mas}$ for scalability. We only changed $\sigma_{AL}$ in Sect.~\ref{section:sigal}, where   we studied the influence of $\sigma_{AL}$ itself. Compared to the $0.1{-}10\unit{mas}$ of $\sigma_{AL}$ \citep{astrometry_solution_edr3}, A19 gives a mean error of 612 mas in the  AC direction by analysing the epoch position error ellipses of the solar objects, which is hundreds of times  the error in AL direction. Thus, we generated the mock data $\eta$ and performed the simulation only in the AL direction. 

We used Eq.~\ref{eq:radial_velocity} to generate the RV data, and added the RV error from a normal distribution,
\begin{equation}
RV(t')_{new}=RV(t')+\mathcal{N}(0,\sigma_{RV})\ (\unit{km\  s^{-1}}),
\end{equation}
where $t'$ is the observation time of the spectrum. 
Gaia can also provide a large amount of  RV data, and the RV error is varied with the apparent magnitude, with a median $\sigma_{RV}$ around $1 \unit{km\ s^{-1}}$.

\section{MCMC simulation}\label{section:MCMC_sim}
In this section we try to obtain a relation between $\Xi$ and the other observable parameters with a series of MCMC simulations. 
First, we introduce the Bayesian model in our simulation in Sect.~\ref{section:bayesian_model}. Then we give the feasible period range in Sect.~\ref{section:feasible_period}. In the following subsections we study the variation of $\Xi$ with other parameters and construct a function, which is called  the $\Xi$-relation for simplicity. 
A19 provide an analytic estimate for $\Xi$,
\begin{equation}\label{eq:and_5yr_KF}
\Xi(a_s,\sigma_{AL},N,N_0)=0.9\cdot(\frac{\sigma_{AL}}{a_s})(\frac{N_0}{N})^{\frac{1}{2}},\ P{<}5\unit{yr},\\
\end{equation}
where $P\!\leq\!5\unit{yr}$ and $N_0=75$. Here $N$ is the number of Gaia observations, while $N_0$ is the number of observations  used in the MCMC simulation. We  not only re-examined the relation between $\Xi$ and $[a_s,\sigma_{AL},N]$, but also explored the relation between $\Xi$ and other parameters, such as $[P,e,i,\beta]$:
\begin{eqnarray}\label{eq:full_relation}
\Xi&\equiv&\Xi(a_s,\sigma_{AL},N,P,e,i,\beta)\nonumber\\
&=&\Phi_0(a_s,P)\cdot\Phi_1(P)\cdot\Phi_2(N,N_0)\nonumber\\
&&\cdot\Phi_3(\sigma_{AL})\cdot\Phi_4(P,e,i)\cdot\Phi_5(\beta).
\end{eqnarray}
In Sects.~\ref{section:main_relation}, \ref{section:obs_num}, and \ref{section:sigal} we study the parameters $[a_s,N,\sigma_{AL}]$ and obtain the functions $\Phi_0$, $\Phi_2$, and $\Phi_3$. In Sect.~\ref{section:inc-ecc} and Sect.~\ref{section:ecliptic}, we derive the functions $\Phi_4$ and $\Phi_5$ for parameters $[e,i,\beta]$. We show the 1yr degeneracy in Sect.~\ref{section:1year} and use the function $\Phi_1$ to avoid this problem.

\subsection{Bayesian model}\label{section:bayesian_model}
For $N$ mock Gaia observations, the chi-square $\chi^2$ is defined by the  equation 
\begin{eqnarray}\label{eq:chisquare}
\chi^2&=&\sum_{i}^{N}(\frac{\eta(t_i)-\eta(t_i)_{\rm{true}}}{\sigma_{AL}})^2,
\end{eqnarray}
where $\eta(t_i)$ is the predicted displacement , while $\eta(t_i)_{\rm{true}}$ is the input mock displacement and $t_i$ is the time of the $i$-th observation. Then the Bayesian posterior probability is the same as Equation 12 in A19, 
\begin{eqnarray}
\ln(posterior)&\sim&\ln(prior)+\ln(likelihood)\nonumber\\
&=&\ln(prior)-\frac{1}{2}\cdot\chi^2,
\end{eqnarray}
where $\ln(prior)$ is the prior of the parameters. A non-informative flat prior is used for $[\Delta\alpha_0\cos\delta, \Delta\delta, \mu_{\alpha}\cos\delta, \mu_{\delta}]$, while uniform distributions with upper and lower limits are used as the prior for the other parameters.

If we have additional $N_{RV}$ mock RVs, then  $\chi^2$ will be
\begin{eqnarray}
\chi^2=\sum_{i}^{N}(\frac{\eta(t_i)-\eta(t_i)_{true}}{\sigma_{AL}})^2+\sum_{j}^{N_{RV}}(\frac{RV(t_j)-RV(t_j)_{true}}{\sigma_{RV}})^2,
\end{eqnarray} 
where $t_j$ is the epoch of the RV. We focus on the astrometry data in this work and give a short discussion on the influence brought by additional RV data.

\subsection{Model feasibility}\label{section:feasible_period}
In order to test the feasibility of the astrometry mass measurement method, we performed simulations for different parameter combinations without adding any observational errors in the position data. From O20 we collected some typical combinations for a BH/NS-LC binary systems. For the case of BH as the primary, $m_p$ is set at $8\,M_{\odot}$, while $m_s=[1,8,50]\,M_{\odot}$. For the case of NS as the primary, $m_p$ is set at $1.4\,M_{\odot}$ and $m_s$ is changed to $[1,5,10]\,M_{\odot}$. The orbit period $P$ is between 2 days and 4500 days. We used short chains here since we started from the true solution to test whether the solution can remain stable around it. For each input source we ran 30 chains for 6500 steps, with 1500 steps as the burn-in. 

\begin{figure}[h]
\centering
\includegraphics[scale=0.6]{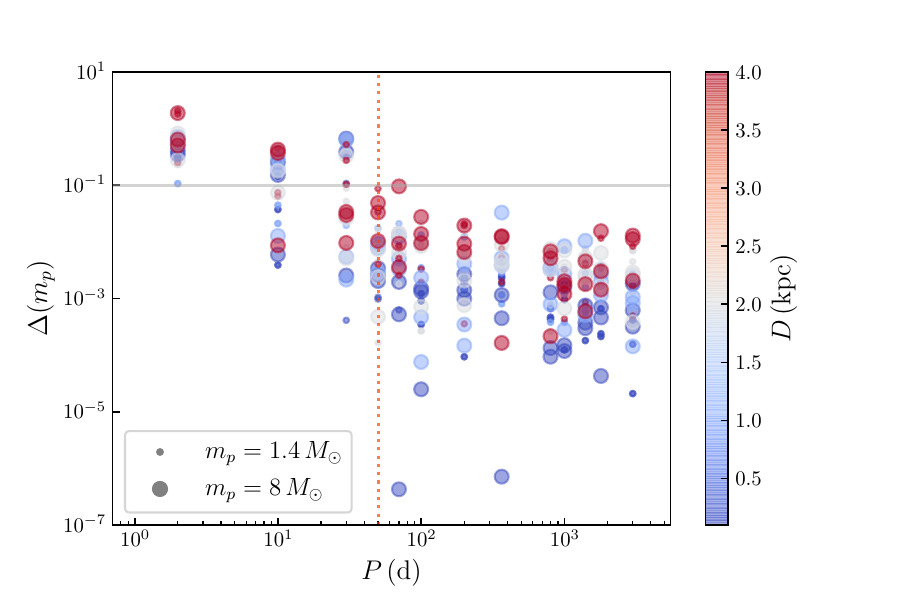}
\caption{Relative bias $\Delta(m_p)$ at different $P$ for the parameter combinations of $[m_p,m_s,P,\varpi]$. The colour of each point stands for a different distance $D$ (see colour bar), which is $1/\varpi$. The size of the point gives the type of the primary, a $1.4\,M_{\odot}$ NS or a $8\,M_{\odot}$ BH. The grey horizontal line indicates  the 0.1 limit, while the red dotted line is $P{=}50\unit{d}$. }
\label{fig:no_error}
\end{figure}

In Fig.~\ref{fig:no_error} we plot $\Delta(m_p)=|m_{p,pre}-m_{p,in}|/{m_{p,in}}$, the relative bias of $m_p$. Here $m_{p,pre}$ is the $m_p$ predicted by simulation result, while $m_{p,in}$ is the input parameter $m_p$. The bias $\Delta(m_p)$ decreases with the input period. When the period is longer than 50 days, $\Delta(m_p)$ is less than $10\%$ of the input $m_p$. For sources with shorter periods $\Delta(m_p)$ can exceed $30\%$, which means the solution might be useless. As a result, we set 50 days as the lower limit of the period of Gaia solvable sources, which is also consistent with the cadence of Gaia \citep{yamaguchi2018,shikauchi2021}. For the sources below this period limit, the observation data may contain more than one unique solution or lead to other incorrect solutions. 

We derived the upper limit using the result from \cite{Lucy2014}, who studied the mass measurement from incomplete astrometric binary orbits. The author points out that the period of a solvable astrometry binary should not be longer than 2.5 times of the observation time in order to keep a small bias in mass measurement. We applied this criterion to a $5\unit{yr}$ Gaia mission, which indicates a $12.5\unit{yr}$ period upper limit. 

\subsection{$\Phi_0$: The main relation of $a_s$ and $P$}\label{section:main_relation}

\begin{figure*}[ht]
\includegraphics[width=1.\textwidth]{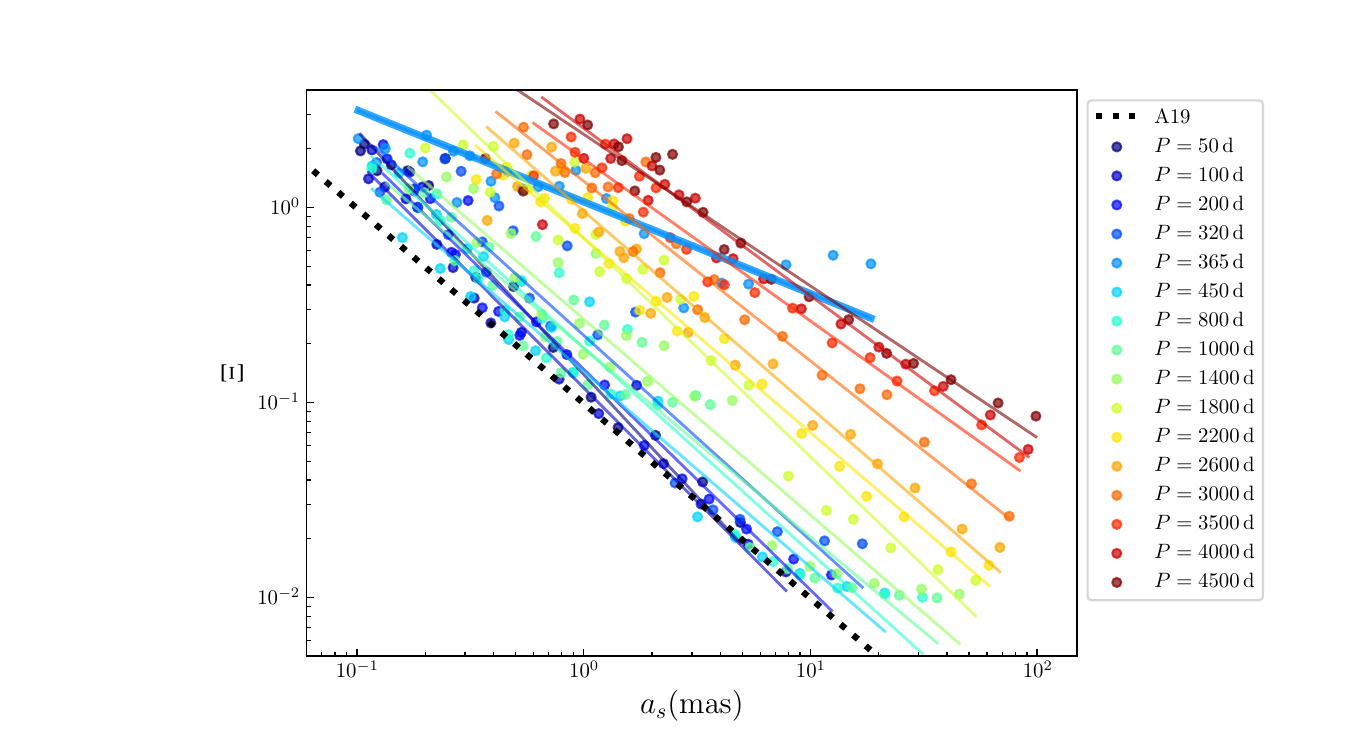}
\caption{ \label{fig:main}Main relation $\Phi_0$ constructed by a set of fitted lines. The simulation results are shown as points of different colours for different periods, from 50 days to 4500 days (see legend at right), while a series of fitted lines in the $\log{-}\log$ space are also plotted in the same colour. The thick blue line is the fitting result of $P{=}1\unit{yr}$. For comparison, the dotted line shows the result of A19, calculated by Eq.~\ref{eq:and_5yr_KF}.}
\end{figure*}

We studied the mass combinations in Sect.~\ref{section:feasible_period}, with $P$ from 50 days to 4,500 days, and thus the semi-major axis $a_s$ varies from $0.009\unit{AU}$ to $9.9\unit{AU}$. Except for $[m_p,m_s,P,a_s]$, the other parameters $[\Delta\alpha_0\cos\delta,\Delta\delta_0,\mu_{\alpha}\cos\delta,\mu_\delta,\varpi,e,i,\omega,\Omega]$ are described in Sect.~\ref{section:mock_data}. The epoch of periastron passage $T_p$ is fixed at $0.3P$ for different $P$. In the simulation of the main relation $\Phi_0$, all of the parameters and their ranges are listed in Table~\ref{table:main_relation_range}. These upper and lower limits are able to cover almost the full range of all the MCMC points. 

\begin{table}      
\centering                          
\begin{tabular}{c c c c}        
\hline\hline                 
Parameter & Typical value & Lower & upper \\    
\hline                        
$\Delta\alpha_0\cos\delta$ &$0.1\unit{mas}$& $-$& $-$  \\
$\Delta\delta_{0}$ &$0.2\unit{mas}$& $-$& $-$  \\
$\mu_{\alpha}\cos\delta$ &$2{\sim}28\unit{mas\cdot yr^{-1}}$& $-$& $-$  \\
$\mu_{\delta}$ &$4{\sim}49\unit{mas\cdot yr^{-1}}$& $-$& $-$  \\
$\varpi$ &$0.25{\sim}10\unit{mas}$& $0.02\unit{mas}$& $4\varpi\unit{mas}$  \\
$a_s$ &$10^{-3}{\sim} 100\unit{mas}$& $10^{-4}\unit{mas}$& $10a_s\unit{mas}$  \\
$P$ &$50d{\sim}12.5\unit{yr}$& $0$ & $10P$ \\
$T_P$& $0.3P$ & $-P$ & $5P$\\
$i$ & $30^{\circ}$ & $0^{\circ}$ & $180^{\circ}$ \\
$\omega$ &$11.5^{\circ}$ &$0^{\circ}$ & $360^{\circ}$ \\
$\Omega$ &$11.5^{\circ}$& $0^{\circ}$ & $180^{\circ}$ \\
$e$ &$0.01$& $0$ & $0.95$\\
\hline                                   
\end{tabular}
\caption{Lower and upper limits of the different paramters used for $\Phi_0$.\label{table:main_relation_range}}
\end{table}

We performed the MCMC simulation six times for each parameter combination. Every time we generated a new set of mock data and ran 30 walkers starting from the neighbourhood around the true parameters. As was done in A19, each sampler walks 40,000 steps, which contains 10,000 steps for burn-in. The walker number and the chain length are also used in the following MCMC simulation of this work. We generated new mock data in each simulation in order to determine Gaia's ability to observe a system rather than a specific data set, which helps us avoid the bias of extreme mock data. We only repeated six times at every point for the purpose of covering the target parameter space efficiently. This method is proved feasible in Sect.~\ref{section:summary_full_relation}.

In Fig.~\ref{fig:main} we used a series of straight lines to fit the $\Xi{-}a_s$ relation with different periods. They have the following forms:
\begin{equation}\label{eq:main_relation}
\log_{10}\Xi=\mathbf{k}\cdot \log_{10}a_s+\mathbf{b}.
\end{equation} 
Therefore, we obtained $\Phi_0(a_s,P){=}\Xi$, which is the main relation of the astrometric mass measurement ability of Gaia. Here, $a_s$ is related to $[P,m_p,m_s,\varpi]$ in the form of 

\begin{equation}\label{eq:a_s}
a_s=\varpi\cdot[\frac{GP^2 m_p^3}{4\pi^2(m_p+m_s)^2}]^{\frac{1}{3}}.
\end{equation} 

\begin{figure}[h!]
\centering
\includegraphics[scale=0.56]{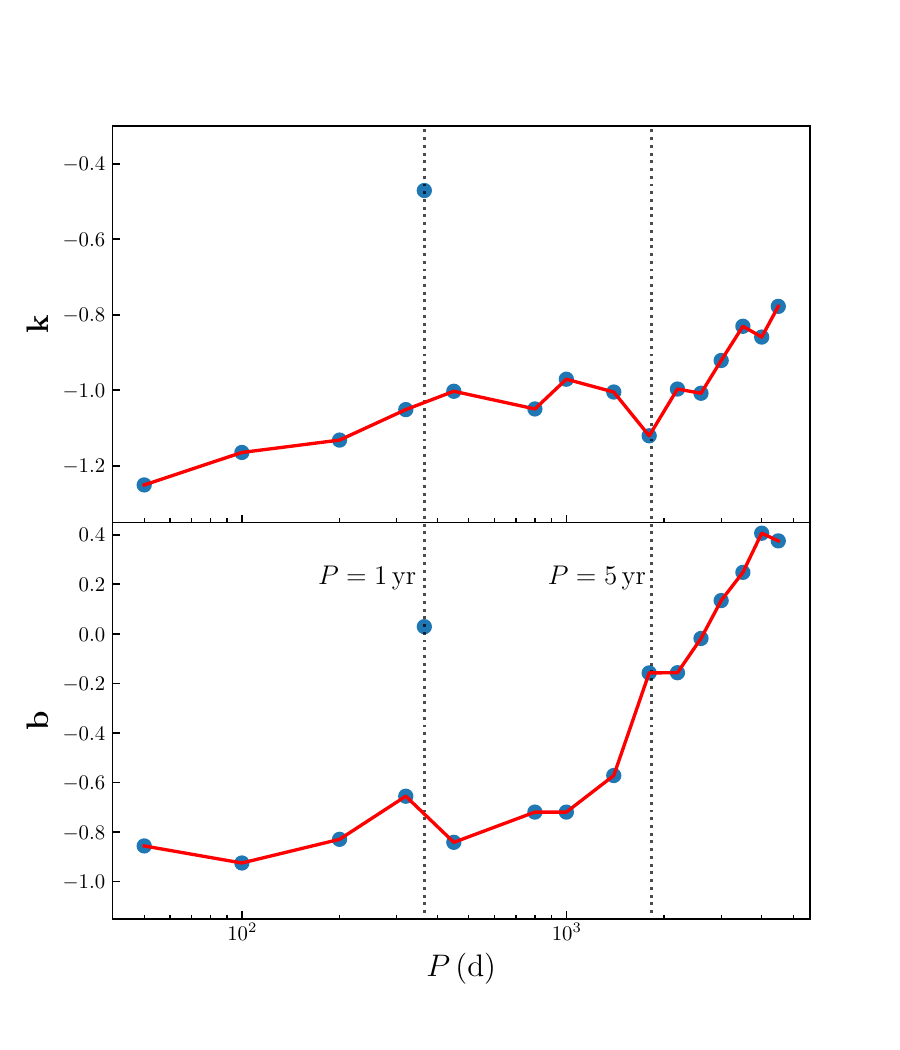}
\caption{Coefficients $\mathbf{k}$ and $\mathbf{b}$ of each line fitted in Fig.~\ref{fig:main}. The dashed grey lines show the periods of $1\unit{yr}$ and $5\unit{yr}$. The outlier is the result at $P=1\unit{yr,}$ which was excluded from  the  study.\label{fig:kb}}
\end{figure}

Combining  the results in Fig.~\ref{fig:kb} and ignoring the special case around 1yr, we find that $\Xi$ is indeed proportional to $a_s^{\mathbf{k}}$ for $P\lqq5\unit{yr}$ with an index $\mathbf{k}$ around -1. The intercept $\mathbf{b}$ is $[-0.8,-0.6]$ for $P\lqq1500\unit{d}$, which is why A19 can use a single line to fit the data within $5\unit{yr}$, while it changes greatly at longer period. In addition, we also find that $\Xi$ around 1 yr shows a specific dispersion due to the degeneracy of the binary orbit and the annual motion. We discuss this phenomenon in Sect.~\ref{section:1year}

There are several possible reasons why our main relation within $5\unit{yr}$ is different from the result in A19 (see dotted line in Fig.~\ref{fig:main}). First, A19 simulated the sources at $[10,100,1000]\unit{pc}$ while we simulated sources at $[10,1,0.5,0.25]\unit{mas}$ corresponding to $[100,1000,2000,4000]\unit{pc}$. Second, we avoided  using an exponential distance prior \citep{Andrews2019,bailer-jones} by using the astrometry observable parallax directly instead of the distance. In order to assign the value of $\sigma_{AL}$, A19 used a $\sigma_{AL}{-}m_G$ relation similar to Figure 9 in \cite{astrometry_solution} which varied with apparent magnitude. This $\sigma_{AL}$ reaches $0.04{-}0.06\unit{mas}$ within $8{-}12.5\unit{mag}$ in the $G$ band, while we used a constant $0.1\unit{mas}$ for all the $\sigma_{AL}$. We only used different $\sigma_{AL}$ in  Sect.~\ref{section:sigal}, which studies the $\sigma_{AL}$ itself. Therefore, our result reflects the true relation between $\Xi$ and $[a_s,P]$, and is easier to apply to different $\sigma_{AL}$ determined by the luminosity of the secondary and the distance.

\subsection{$\Phi_1$: Degeneracy around one year}\label{section:1year}
\begin{figure}[h!]
\centering
\includegraphics[scale=0.6]{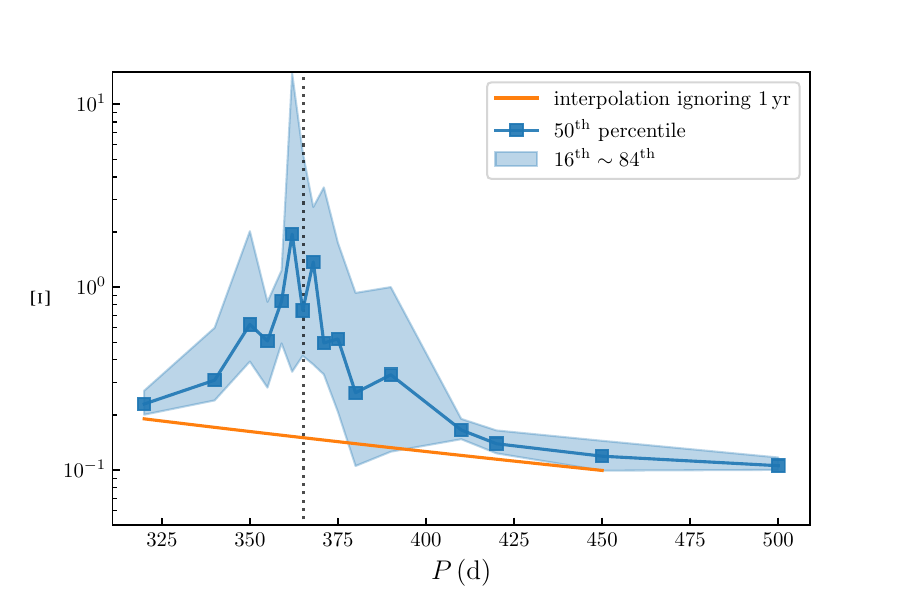}
\caption{Value of $\Xi$ affected by the 1\unit{yr} degeneracy. The blue squares are the median values  of the 50  simulations of each period. The blue shadow is the area between the  $16{\rm{th}}$ percentile and the  $84{\rm{th}}$ percentile. The orange line is calculated with Eq.~\ref{eq:main_relation}, with $\mathbf{k}$ and $\mathbf{b}$ obtained from the red line in Fig.~\ref{fig:kb}, ignoring the outlier of $P{=}1\unit{yr}$.}
\label{fig:1yr}
\end{figure}
In Fig.~\ref{fig:main} the line at $1\unit{yr}$ shows an outlier of these fitting lines. The $\Xi$ becomes quite large due to the coupling of the annual motion of the Earth and the $1\unit{yr}$ period orbit of the source. A similar phenomenon,  shown by \cite{Holl2011}, is that the detectability of the astrometric planet with Gaia decreased at around 1 yr. In Fig.~\ref{fig:1yr} we chose a system with $[m_p,m_s,\varpi]$ as $[8\,M_{\odot},8\,M_{\odot},1\unit{mas}]$, and varied the period around $1\unit{yr}$. We repeated each simulation ten times in order to get the uncertainty of $\Xi$ at the same time.

In Fig.~\ref{fig:1yr} there is a clear peak around 1 yr. The simulation result is close to the interpolation model at $P{=}320\unit{d}$ and $P{=}420\unit{d}$, which is calculated by Eq.~\ref{eq:main_relation} with  $\mathbf{k}$ and $\mathbf{b}$ obtained from the red line in Fig.~\ref{fig:kb}. To avoid the influence from the 1yr degeneracy, we just omitted the sources with periods within $320\unit{d}$ and $420\unit{d}$ by 
\begin{equation}\label{eq:phi_1yr}
\Phi_1(P)=0,\ 320\unit{d}\lqq P\lqq420\unit{d}.
\end{equation}

We did this for two reasons. First, $\Xi$ around $1\unit{yr}$ is enlarged by several times due to the $1\unit{yr}$ degeneracy; only a few sources in this period range would satisfy our solvable criteria in Sect.~\ref{section:Gaia_solvable_population}. Second, we find that only $3\%$ of the BH/NS-LC systems are $P\in[320\unit{d},420\unit{d}]$ in our simulated sample in Sect.~\ref{section:Gaia_solvable_population}. Since the 1yr degeneracy has limited influence on our final BH/NS-LC population, we just omitted this period range. 

\subsection{$\Phi_2$: Observation number $N$}\label{section:obs_num}
Gaia shows a strong scanning pattern in observation numbers, astrometric errors, among  others; therefore, the sources at different places usually have different results in observation numbers and scanning angle distributions \citep[e.g.][]{astrometry_solution,astrometry_solution_edr3}. For example, we have 62 observations at the location of LB-1 from \textsf{GOST}, which is different from the observation information used in A19. Here we ran the simulations for all the parameter combinations in Sect.~\ref{section:main_relation} with 124 observation epochs, which is twice the number of epochs at this position. Instead of the original epochs and angles in the \textsf{GOST} file, we added an additional epoch for each epoch with the same scanning angle. The additional epoch is sampled from a Gaussian distribution,  $\mathcal{N}(0\unit{d},20\unit{d})$. We calculate the value of $\Xi_{N=124}/\Xi_{N=62}$, which has a median ratio $0.70^{+0.29}_{-0.18}$. 

If the mission extended to ten years rather than five years, the longer observation time would also accumulate more observations. Thus, \cite{astrometry_solution_edr3} gives an expected improvement scale as $T^{-1/2}$ for the parallaxes, positions, and their uncertainty. Here we extrapolated this  scale to our mass measurement $\Xi$:
\begin{equation}
\frac{\Xi_{T}}{\Xi_{T=T_0}}=(\frac{T}{T_0})^{-\frac{1}{2}}, \ (T_0=5\unit{yr}). \label{eq:T_T0}
\end{equation}
If extending the observation mission were the same as increasing the observation number, Eq.~\ref{eq:T_T0} would be the same as the following equation:
\begin{equation}
\Phi_2(N,N_0)=(\frac{N}{N_0})^{-\frac{1}{2}},\ (N_0=62) \label{eq:N_N0}.
\end{equation}
Our scale is $0.70^{+0.29}_{-0.18}$, very close to   $1/\sqrt{2}$ or $0.707$. Thus, we confirm the scale of Eq.~\ref{eq:N_N0}.
For simplicity,
we only considered the variation caused by increasing the number of epochs for a $10\unit{yr}$ Gaia mission in this work. If we do a simulation up to $25\unit{yr}$ according to the analysis in Sect.~\ref{section:feasible_period}, we would get a more realistic results for a $10\unit{yr}$ Gaia mission.

\subsection{$\Phi_3$: Gaia epoch data uncertainty $\sigma_{AL}$}\label{section:sigal}
The uncertainty of Gaia epoch data $\sigma_{AL}$ is a direct variable that affects the mass measurement, which appears in Eq.~\ref{eq:and_5yr_KF}: $\Xi{\sim}\frac{a_s}{\sigma_{AL}}$. In the study of astrometric planets, \cite{perryman2014} defined the astrometric signal-to-noise ratio, $S/N{\equiv}\frac{\alpha}{\sigma_{FOV}}{=}\frac{a_s}{\sigma_{AL}}$ (the form in this work). We used a constant $\sigma_{AL}$ in most of our simulations; therefore, we decided to study this parameter in an additional simulation test, which is equivalent to studying the effect of the $SNR$. 

\begin{figure}[h!]
\centering
\includegraphics[scale=0.6]{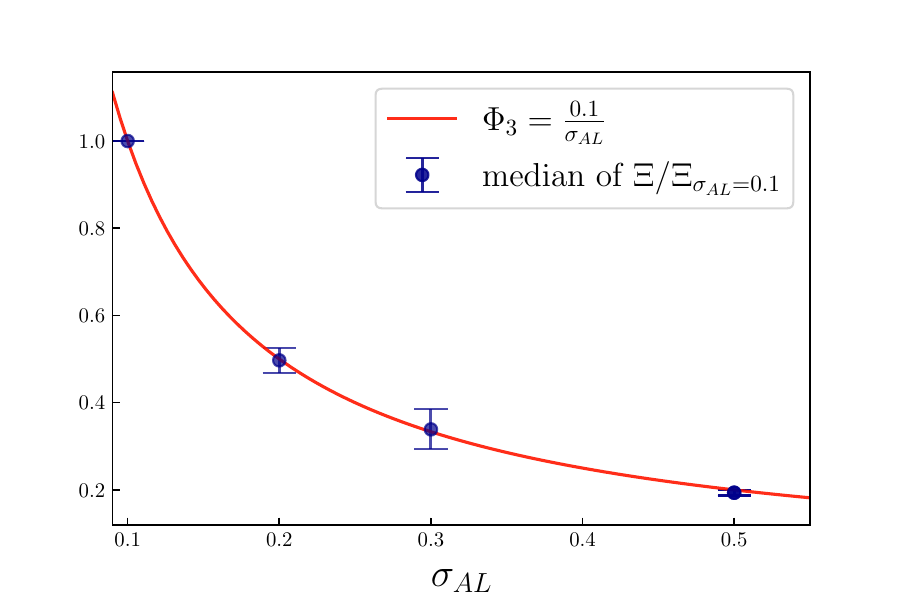}
\caption{Values of $\Xi$ normalised to $\Xi_{\sigma_{AL}=0.1}$. The blue points ares the median of all the values of $\Xi/\Xi_{\sigma_{AL}=0.1}$ with the same $\sigma_{AL}$. The red line is the $\Phi_3$ that equals $\frac{0.1}{\sigma_{AL}}$.}
\label{fig:sigal}
\end{figure}

We studied a BH-LC system with $[m_p,m_s]$ as $[8\,M_{\odot},8\,M_{\odot}]$, where $a_s$ is proportional to $\varpi\cdot P^{\frac{2}{3}}$. So we chose different periods for sources at $[0.5,1,10]\unit{mas}$, of which the $SNR$ falls in the range [1,100]. In Fig.~\ref{fig:sigal} we normalised the $\Xi$ to $\Xi(\sigma_{AL}{=}0.1)$, and we got a quite good relation of $\Phi_3(\sigma_{AL})$:
\begin{eqnarray}\label{eq:phi_sigal}
\Phi_3(\sigma_{AL})=\frac{0.1}{\sigma_{AL}}.
\end{eqnarray}
It should be noted  that only the sources with $\varpi\rqq\sigma_{AL}$ and $a_s\rqq1.5\sigma_{AL}$ follow this relation. The sources with smaller $\varpi$ or $a_s$ would get higher $\Xi$. 

\subsection{$\Phi_4$: Inclination $i$ and eccentricity $e$}\label{section:inc-ecc}
\begin{figure*}[ht]
\centering
\includegraphics[scale=0.7]{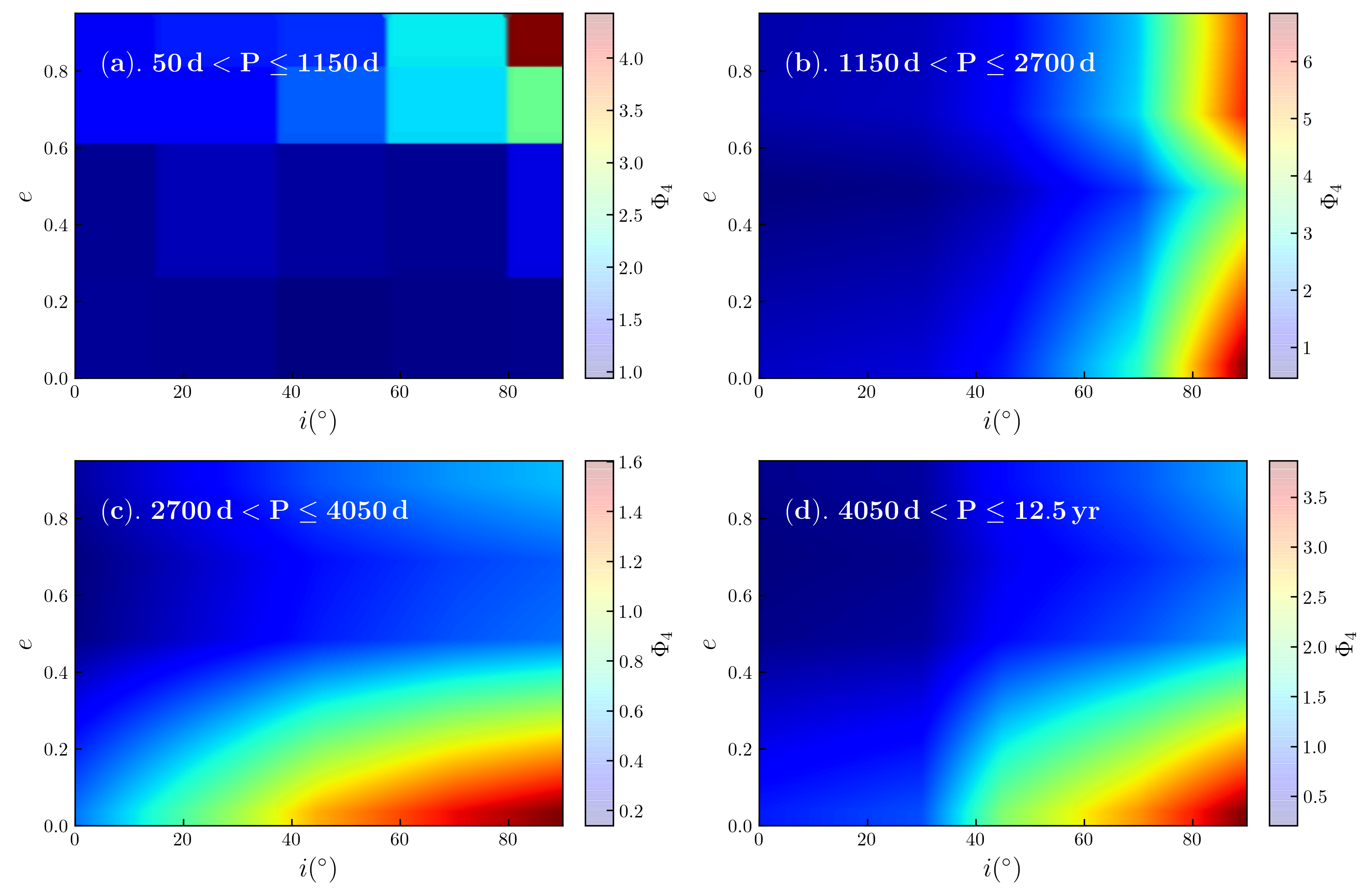}
\caption{Values of $\Phi_4$ varied with $i$ and $e$ in different period ranges. For $50\unit{d}{<}P\leq1150\unit{d}$  a 2D nearest-neighbour interpolation function is used to fit $\Phi_4$. When $P$ falls in the range of [1150\unit{d},2700\unit{d}], [2700\unit{d},4050\unit{d}] and [4050\unit{d},12.5\unit{yr}],  linear interpolation functions are used to fit $\Phi_4$. \label{fig:example_ecc_inc}}
\end{figure*}

In this section we describe the simulations we performed for different combinations of $[i,e]$ when $P$ is $[500,1825,3650,4500]\unit{d}$. For each period $P$, we fixed the $[m_p,m_s,\varpi]$ at $[8\,M_{\odot},8\,M_{\odot},1\unit{mas}]$. The other parameters are the same as in Sect.~\ref{section:main_relation}, except for the inclination angle $i$ and the eccentricity $e$. We simulated six times for every combination of $[i,e]$ when $i{=}[0^{\circ},30^{\circ},45^{\circ},70^{\circ},90^{\circ}]$ and $e{=}[0.01,0.5,0.6,0.75,0.90]$.

We normalised the $\Xi$ of each point to $\Xi(e=0.01,i=30^{\circ})$ and find that it is hard to give a simple equation for $\Phi_4(P,e,i)$. Therefore, we divided the period into four parts and use interpolation functions to describe $\Phi_4$ (see Fig.~\ref{fig:example_ecc_inc}). 

There is a total trend in the four panels of Fig.~\ref{fig:example_ecc_inc} that $\Phi_4$ increases   when  $i$ gets larger. At high inclination, $\Phi_4$ increases to at high eccentricity in panel \textbf{a}, while it has an opposite relation in panels \textbf{c} and \textbf{d}. In panel \textbf{b}, $\Phi_4$ increases   at both the higher and lower end of $e$, and decreases   in the middle. The ratio of the maximum to the  minimum of $\Phi_4$ in the four panels could reach 4.5 to 8. Thus, $\Phi_4$ would influence the final distribution of eccentricity, which should not be omitted in statistical studies.

\subsection{$\Phi_5$: Ecliptic latitude $\beta$}\label{section:ecliptic}
\begin{figure}[h!]
\centering
\includegraphics[scale=0.6]{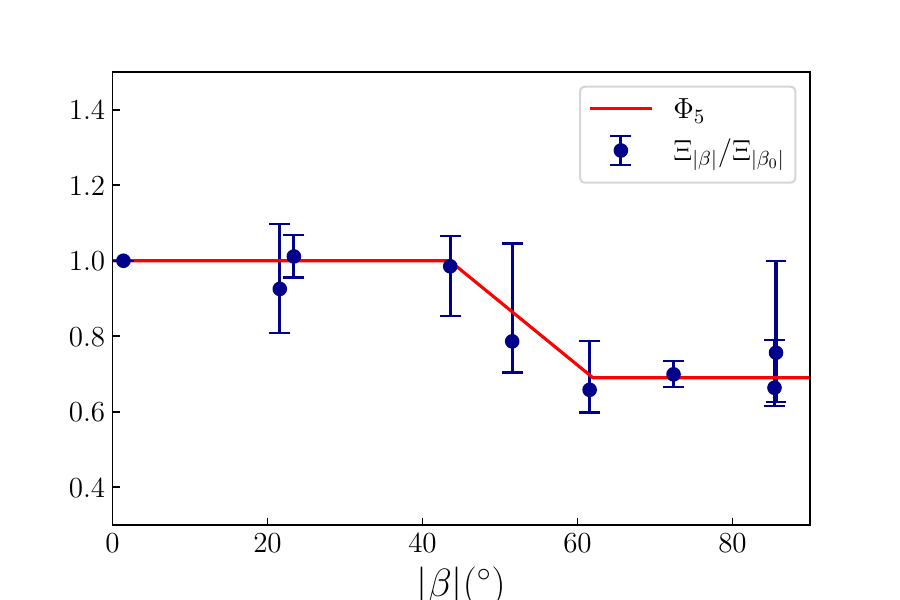}
\caption{$\Phi_5$ fitted to the ratio of $\Xi_{|\beta|}$ and $\Xi_{|\beta_0|}$. Each point is the median of the ratio for all the periods in [100\unit{d},4500\unit{d}] with the same $|\beta|$. The $\Xi$ drops to $0.7\Xi_{|\beta_0|}$ when $|\beta|$ is larger than $62^{\circ}$. }
\label{fig:dec}
\end{figure}

Considering in addition a $[8\,M_{\odot},8\,M_{\odot}]$ system at $1\unit{mas}$, we select nine positions for the simulations such that the absolute value of their ecliptic latitude $|\beta|$  covers the range $[0^{\circ},85^{\circ}]$. The period falls in the range of [100d,4500d], while the other parameters are the same as in Sect.\ref{section:main_relation}. In order to perform  simulations at different positions, we used new observation data from the \textsf{GOST} program of each position.

In Fig.~\ref{fig:dec} we show the ratio $\Xi/\Xi_{|\beta|=1.4^{\circ}}$ versus $|\beta|$. All of the ratio values have been corrected with observation numbers according to Sect.~\ref{section:obs_num}. The $\Xi$ is nearly the same as $\Xi_{|\beta|=1.4^{\circ}}$ from $|\beta|{=}1.4^{\circ}$ to $|\beta|{=}43^{\circ}$, while it drops to $0.7\Xi_{|\beta|=1.4^{\circ}}$ over $|\beta|{=}62^{\circ}$. Here the point with $\beta{=}1.4^{\circ}$ corresponds to the position $[\alpha_0,\delta_0]{=}[92.95^{\circ},22.82^{\circ}]$. Thus, $\Phi_5(\beta)$ is a piecewise function:
\begin{eqnarray}
\Phi_5(\beta)=\left\{\begin{array}{ccc}1\ ,&|\beta|\leq43^{\circ}\\ 
-0.0158|\beta|+1.679, &43^{\circ}\lqq|\beta|\leq62^{\circ}\\
0.7, &|\beta|\rqq62^{\circ}\end{array}. \right.
\end{eqnarray}

\subsection{Radial velocity}\label{section:radial_velocity}

In this section we focus on a NS-LC system located at $[0.5,1,10]\unit{mas}$ which contains a $1.4\,M_{\odot}$ NS and a $1\,M_{\odot}$ companion. We used the same parameters in Sect.~\ref{section:main_relation} with 17 additional  RVs,  and repeated the simulation six times for each $P$. The epoch of RV observation is selected from the \textsf{GOST} file randomly in each simulation for simplicity and universality. The RV uncertainty is a Gaussian error with $\sigma_{RV}=1\unit{km\ s^{-1}}$.

\begin{figure}[h!]
\centering
\includegraphics[scale=0.6]{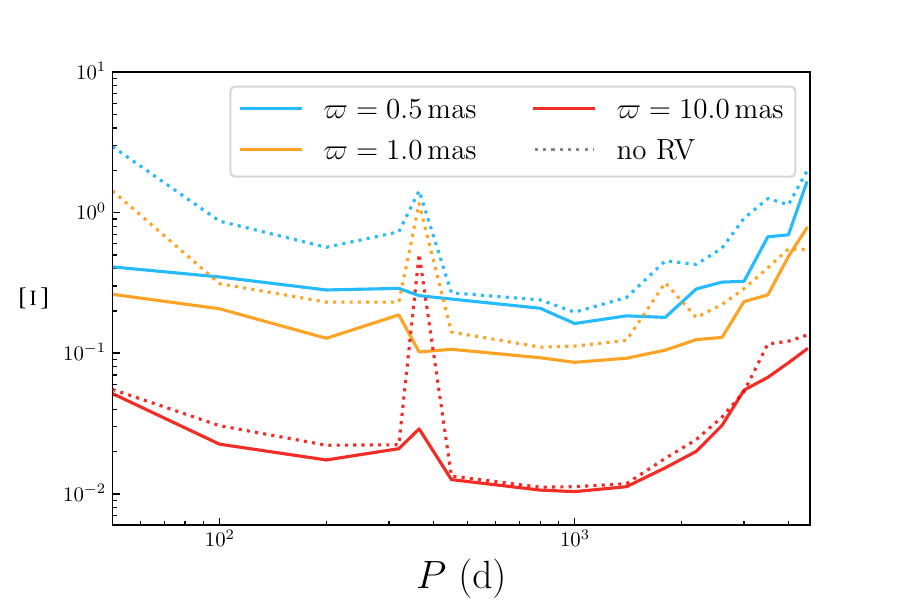}
\caption{Values of $\Xi$ of a NS-dwarf system ($1.4\,M_{\odot}\!-\!1\,M_{\odot}$) with inclination angle $i\!=\!30^{\circ}$. Sources at $[10,1 ,0.5]\unit{mas}$ are plotted in red, yellow, and blue, respectively. As a comparison,  the dashed line is plotted in  the same colour as the $\Xi$ of the same source with no RV data. }
\label{fig:with_rv}
\end{figure}

In Fig.~\ref{fig:with_rv} it is clear that the additional RV data could improve the solution for sources at $[0.5,1]\unit{mas}$, while it helps little for sources at $10\unit{mas}$. Considering Eq.~\ref{eq:kepler_factor}, $\Xi$ is proportional to $\frac{a_s^3}{\varpi^3P^2}$. Therefore, the RV data cannot help constrain the parameters better, such as $[a_s,\varpi,P]$, if they have already reached the astrometry limit.  Furthermore, the effect of RV data are not as significant at larger $a_s$ as at smaller $a_s$. The reason is that the smaller $a_s$ corresponds to a shorter period, which leads to a higher $\frac{K}{\sigma_{RV}}$. The greatest improvement happens at $P{=}1\unit{yr}$, showing that the RV decouples the 1yr degeneracy of $a_s$ and $\varpi$. After checking the uncertainty of many parameters or parameter combinations, we find that at $P{=}1\unit{yr}$ the RV helps constrain the uncertainty of orbit radius $R_s\,({\equiv}\frac{a_s}{\varpi})$ down to $5\%$ of the same value   calculated from the simulation without RV data. These phenomena suggest that RV is very useful to get a better binary solution. 

\subsection{Summary of the $\Xi$-relation}\label{section:summary_full_relation}

The final result of our simulation, the $\Xi$-relation, is a function containing six subfunctions. In summary, $\Phi_0(a_s,P)$ and $\Phi_1(P)$ describe the overall profile of the full relation; $\Phi_2(N)$ and $\Phi_3(\sigma_{AL})$ give us the result of the variation in the  Gaia observations; and   $\Phi_4(P,e,i)$ and $\Phi_5(\beta)$ tell the influence of the orbit shape and the system location. The epoch observation error $\sigma_{AL}$ is determined by the apparent magnitude $m_G$ from   Fig. A.1 in \cite{astrometry_solution_edr3}. Therefore, we limit the $m_G$ within $[6^m,21^m]$. In fact, most sources in Gaia would be fainter than $6^m$. The sources brighter than $6^m$ might have worse epoch precision according to the same figure. 
The domain of our function is summarised in Table ~\ref{tab:domain_table}.

\begin{table}[h!]
\centering                          
\begin{tabular}{c c }        
\hline\hline                 
Parameter & Domain\\    
\hline\hline
$\sigma_{AL}$ &$\varpi>\sigma_{AL},\ a_s>1.5\sigma_{AL}$\\
$P$ & $50\unit{d}{<}P{<}4565\unit{d}\,(12.5\unit{yr})$\\
$m_G$ & $6^m{<}m_G{<}21^m$\\
\hline                                   
\end{tabular}
\caption{Domain of the $\Xi$-relation.\label{tab:domain_table}}
\end{table}

\begin{figure}[h]
\centering
\includegraphics[scale=0.6]{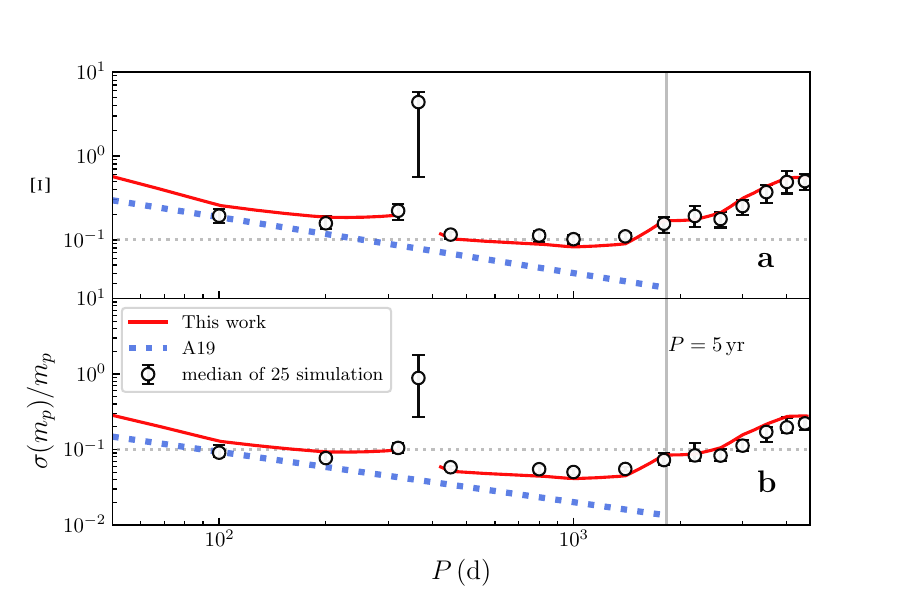}
\caption{One example of a simulated source and the $\Xi$-relation. \textbf{a}: Median of $\Xi$ (black points) calculated from the simulation result in Sect.~\ref{section:summary_full_relation} and the prediction of our model  (red line) and the  A19 model (blue dashed line). \textbf{b}: Median of $\frac{\sigma(m_p)}{m_p}$ (black points) obtained from the simulation result directly, and the predicted $\frac{\sigma(m_p)}{m_p}$ by applying Eq.~\ref{eq:mass_error_propagation} to our predicted $\Xi$  (red line) and that of  A19 (blue dashed line). In both panels the grey vertical line indicates   the $5\unit{yr}$ limit in the  A19 model, and the grey horizontal dashed line shows the difference in $\Xi$ and $\frac{\sigma(m_p)}{m_p}$. The interruption of our model around 1yr is determined by $\Phi_1(P)$.}
\label{fig:error}
\end{figure}

To show the ability of the $\Xi$-relation, we give an example with fixed $[m_p,m_s,\varpi]$ at $[8\,M_{\odot},8\,M_{\odot},1\unit{mas}]$ and varied period in the range of $[100\unit{d},4500\unit{d}]$. We repeated the simulation 25 times for each parameter combination to estimate the uncertainty of $\Xi$. Figure \ref{fig:error}.\textbf{a} not only  gives the value of $\Xi$, but it also tells us the uncertainty of this error. Since we omitted the period around $1\unit{yr}$, most uncertainty of $\Xi$ is below 1\%, which means that it is feasible for us to get the $\Xi$-relation with only six repeats of each simulated data point  mentioned in Sect.~\ref{section:main_relation}. 

In Fig.~\ref{fig:error}.\textbf{a} we plot the predicted $\Xi$ from our final relation. The $\Xi$-relation is almost consistent with the simulation except for a tiny difference at $P{<}100\unit{d}$. Figure \ref{fig:error}.\textbf{b} shows the $\frac{\sigma(m_p)}{m_p}$, calculated from $\Xi$ using Eq.~\ref{eq:mass_error_propagation}, which is also consistent with the point obtained directly from the simulation. This suggests the correctness of Eq.~\ref{eq:mass_error_propagation}, which is very useful in studying other astrophysical problems. Here we simply set $\sigma(m_s)$ equal to  0. Compared to the uniform model in A19, our $\Xi$-relation  extends to longer periods (up to $12.5\unit{yr}$), and it is also more adaptable when the period is shorter than $5\unit{yr}$.

\section{Application of the $\Xi$-relation }\label{section:Gaia_solvable_population} 
In this section we give two cases of application of the $\Xi$-relation. We introduce the method to obtain the apparent magnitude $m_G$ at the beginning. Then in the first case we show the solvable area of some specific systems. In the second case we applied the $\Xi$-relation to a mock Galactic BH/NS-LC population to predict the number of Gaia solvable BH/NS-LCs. In both case we use Eq.~\ref{eq:mass_error_propagation} to calculate $\frac{\sigma(m_p)}{m_p}$ from $\Xi$, in which we assume $\sigma(m_s)$, the error of $m_s$, is $0.1m_s$ obtained by other method.

\subsection{Apparent magnitude}\label{section:apparent_magnitude}
To obtain  $\sigma_{AL}$ we used the apparent magnitude $m_G$ to interpolate from Fig. A.1 in \cite{astrometry_solution_edr3}. Given that the absolute magnitude of the secondary is $M_G$, the apparent magnitude $m_G$ is determined by
\begin{eqnarray}
m_G=M_G+5\cdot \log_{10}\frac{D}{10\unit{pc}}+A_G,
\label{eq:apparent_magnitude}
\end{eqnarray}
where $D$ is the distance to our Solar System and $A_G$ is the interstellar extinction. We use a dust plane model to calculate this extinction \citep[e.g.][]{dust_disk0,dust_disk1},
\begin{eqnarray}
A_V&=&\rho_0\int^{D}_0 \exp(-\frac{r|\sin(b)|}{H})\,dr\nonumber\\
&=&\left\{\begin{array}{ll}\ \ \rho_0\times D\ ,&b=0^{\circ}\\ 
\frac{\rho_0 H}{|\sin(b)|}(1-\exp(-\frac{D\cdot |\sin(b)|}{H})), &b\neq0^{\circ} \end{array},\right.
\end{eqnarray}
where $b$ is the source's Galactic latitude. The scale height of the Galactic dust plane model in \cite{dust_disk1} is  $H{=}164\unit{pc}$. This small scale height means that the dust concentrate around the Galactic plane, and the source with higher Galactic latitude would be less influenced by the extinction. The mean density of the dust plane, estimated by \cite{Chen2016} using 2000 nearby open clusters, is $\rho_0{=}\unit{0.54}\unit{mag\,kpc^{-1}}$. We use the relative extinction value $A_G/A_V{=}0.789$ \citep{wangshu2019} to convert the $V$-band extinction $A_V$ to $A_G$ in Gaia's $G$ band. Finally, the $\sigma_{AL}$ could be obtained from Fig. A.1 in \cite{astrometry_solution_edr3} with $m_G$.

\subsection{Gaia's ability to solve systems with a dark companian}\label{section:ability}

We test the ability of Gaia to measure the mass of the dark component in a binary as an example of using the $\Xi$-relation. This dark component could   be a BH or NS, or even a WD or a planet. Here, we set $\frac{\sigma(m_p)}{m_p}{<}30\%$ as the criterion for being astrometrically solvable.
\begin{figure}[h!]
\centering
\includegraphics[scale=0.25]{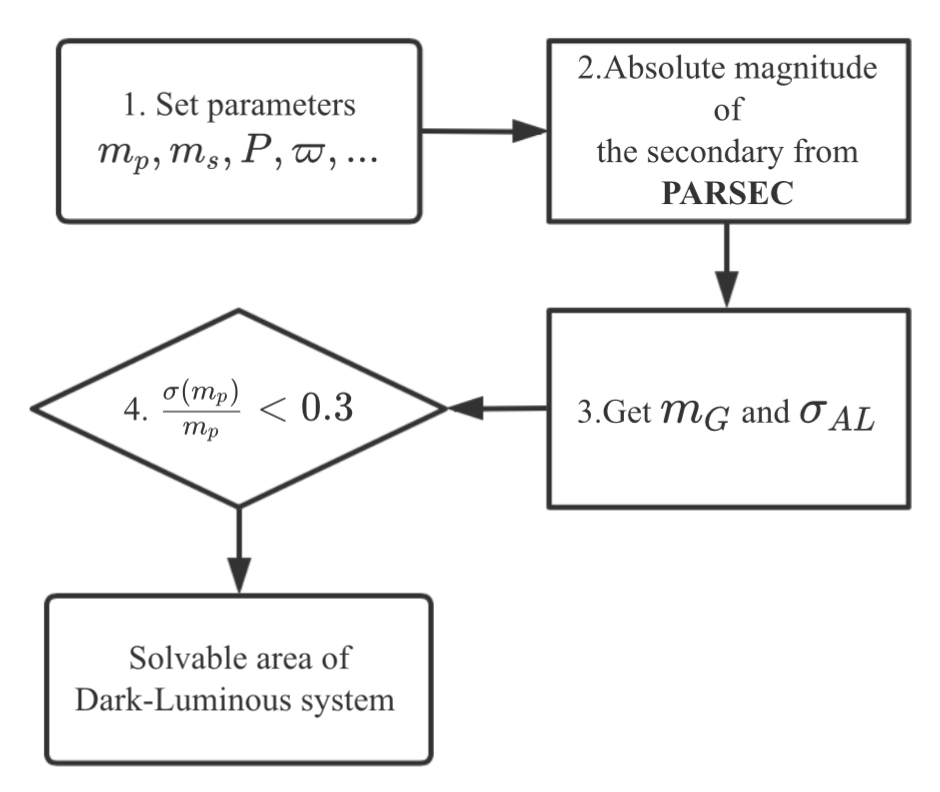}
\caption{Process to explore the Gaia solvable area for the dark-luminous systems in Sect.~\ref{section:ability}.}
\label{fig:process_solvable}
\end{figure}

Figure \ref{fig:process_solvable} gives the process to explore the solvable area for different systems. In Step 1 we set the system parameters; in particular, we varied the system period in $50\unit{d}{\sim}12.5\unit{yr}$ and the parallax in $0.05{\sim}100\unit{mas}$. In Step 2 the typical absolute magnitude in $G$ band of the secondary are derived from the  PARSEC model \citep{CMD}. Then, the apparent magnitude $m_G$ and  $\sigma_{AL}$ are obtained according to Sect.~\ref{section:apparent_magnitude} in Step 3. After checking the solvable criteria in Step 4, it is able to give the solvable area for such a dark-luminous system. In this section the location is fixed at $[\alpha_0,\delta_0]$, which is the location of LB-1. Although the $\Xi$-relation works well for nearby sources, we still set $10\unit{pc}$ as the minimum distance for it. Since our model describes the motion of the source in a tangent plane, it might not hold for sources too close to the Solar System.

\begin{figure*}[ht]
\includegraphics[scale=0.5]{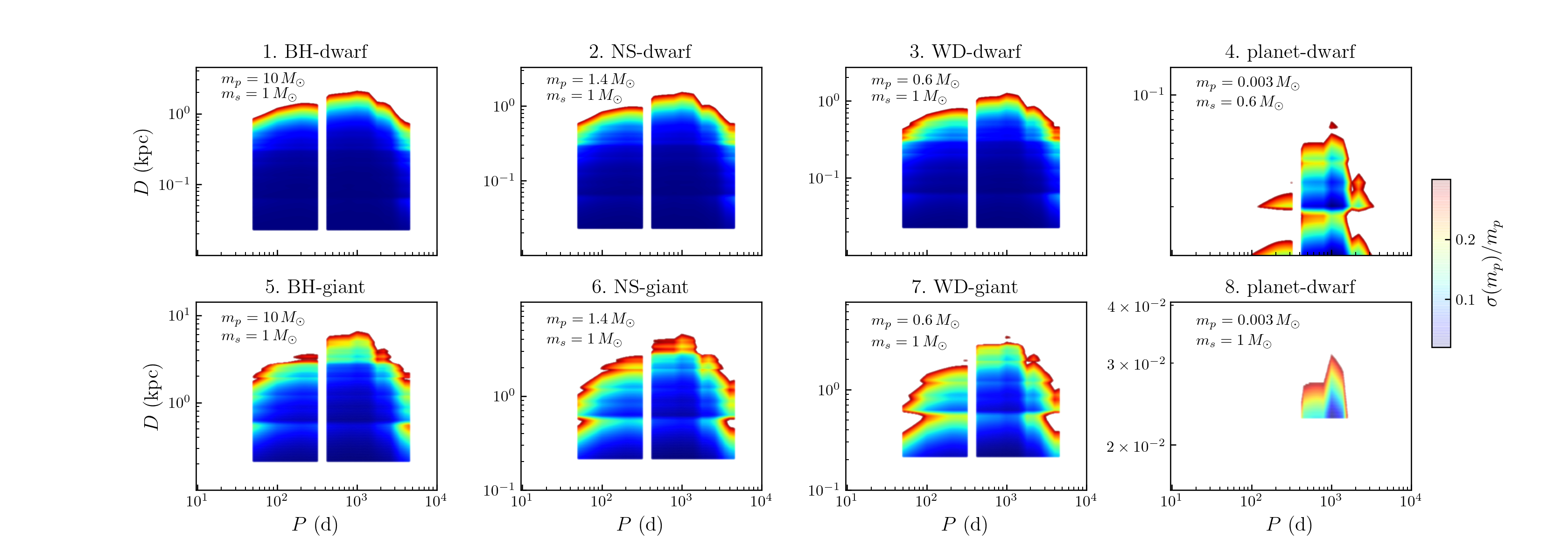}
\caption{Solvable domain for eight combinations of primary and secondary. The parameters and results are listed in Table~\ref{tab:observe_domain}. The luminous companion in panels $[1,2,3,4,8]$ is a dwarf, and is a giant in panels $[5,6,7]$. The colour stands for the value of $\sigma(m_p)/m_p$, from 0 to 0.3. Different y limits are used in each panel, as indicated by ticks on the y-axis. \label{fig:observe_domain}}
\end{figure*}

\begin{table}
\centering                          
\begin{tabular}{c c c c}        
\hline\hline                 
Type of Binary & $m_p,m_s$ & solvable range & Panel number \\
& $(M_{\odot})$ & & in Fig.~\ref{fig:observe_domain}\\[0.5ex] 
\hline\hline                     
BH-dwarf &$10,1$&$0.02{-}2.06\unit{kpc}$&1\\
NS-dwarf &$1.4,1$&$0.02{-}1.51\unit{kpc}$&2\\
WD-dwarf &$0.6,1$&$0.02{-}1.24\unit{kpc}$&3\\
BH-giant &$10,1$&$0.21{-}6.44\unit{kpc}$&5\\
NS-giant &$1.4,1$&$0.22{-}4.54\unit{kpc}$&6\\
WD-giant &$0.6,1$&$0.22{-}3.40\unit{kpc}$&7\\
planet-dwarf &$0.003,0.6$&$10{-}68\unit{pc}$&4\\
planet-dwarf &$0.003,1$&$23{-}31\unit{pc}$&8\\
\hline                                   
\end{tabular}
\caption{Solvable range for different system types in Sect.~\ref{section:ability} and the panel number in Fig.~\ref{fig:observe_domain}. \label{tab:observe_domain}}
\end{table}

We calculated the value $\frac{\sigma(m_p)}{m_p}$, with Eq.~\ref{eq:mass_error_propagation}, for different $[m_p,m_s]$ combinations in Table~\ref{tab:observe_domain}. For a BH/NS/WD-LC system, if the luminous component is a $1\,M_{\odot}$ dwarf, the solvable area would be from 20pc to $2.06, 1.51$, and $1.24\unit{kpc}$ for BHs, NSs, and WDs, respectively. On the other hand, if the luminous component is a $1\,M_{\odot}$ giant, this area would be $0.21\unit{kpc}$ to $6.44, 4.54$, and $3.40\unit{kpc, respectively}$. The farthest limit of this solvable area is determined by the source with $P{\sim}1200\unit{d}$, while the nearest limit is determined by the bright end of the $\sigma_{AL}{-}m_{G}$ relation. 
For exoplanets we study a $0.003\,M_{\odot}$ planet, about $3.2\,M_{Jup}$, which orbits a $1\,M_{\odot}$ dwarf or a $0.6\,M_{\odot}$ dwarf. For consistency, we still use $m_s$ for the luminous star and $m_p$ for the planet. We find that Gaia is able to do the mass measurement of such a planet around the $0.6\,M_{\odot}$ dwarf in $68\unit{pc}$ with a precision better than $30\%$. The solvable area of a planet around $1\,M_{\odot}$ dwarf shrinks to only $31\unit{pc}$ because the astrometric wobble of the dwarf caused by the planet is smaller than the wobble of a $0.6\,M_{\odot}$ dwarf. We didn't run the simulation for a giant with a planet, since the nearest limit of a giant with BH, NS, and WD is $220\unit{pc}$, which is much farther than the upper limit of a planet-LC system.

In Fig.~\ref{fig:observe_domain} we show the astrometrically solvable domain for different systems. These plots have the following features in common: 1) the mountain-like profile,  determined by our $\Xi$-relation; 2) a gap around 1yr  caused by the 1yr degeneracy; and  3)  a clear lower distance limit for a bright secondary. In addition, taking   panel 6 in Fig.~\ref{fig:observe_domain} as an example, when the period is shorter than $1\unit{yr}$, the precision is worse in the middle. Then it gets better at some distance, and finally  goes  to $30\%$. This feature is very clear in panels 4-7, which is caused by the non-monotonic $\sigma_{AL}{-}m_{G}$ relation. 

\subsection{The Gaia solvable BH/NS-LC population}\label{section:Gaia_solvable_population}
\begin{figure}[h!]
\centering
\includegraphics[scale=0.265]{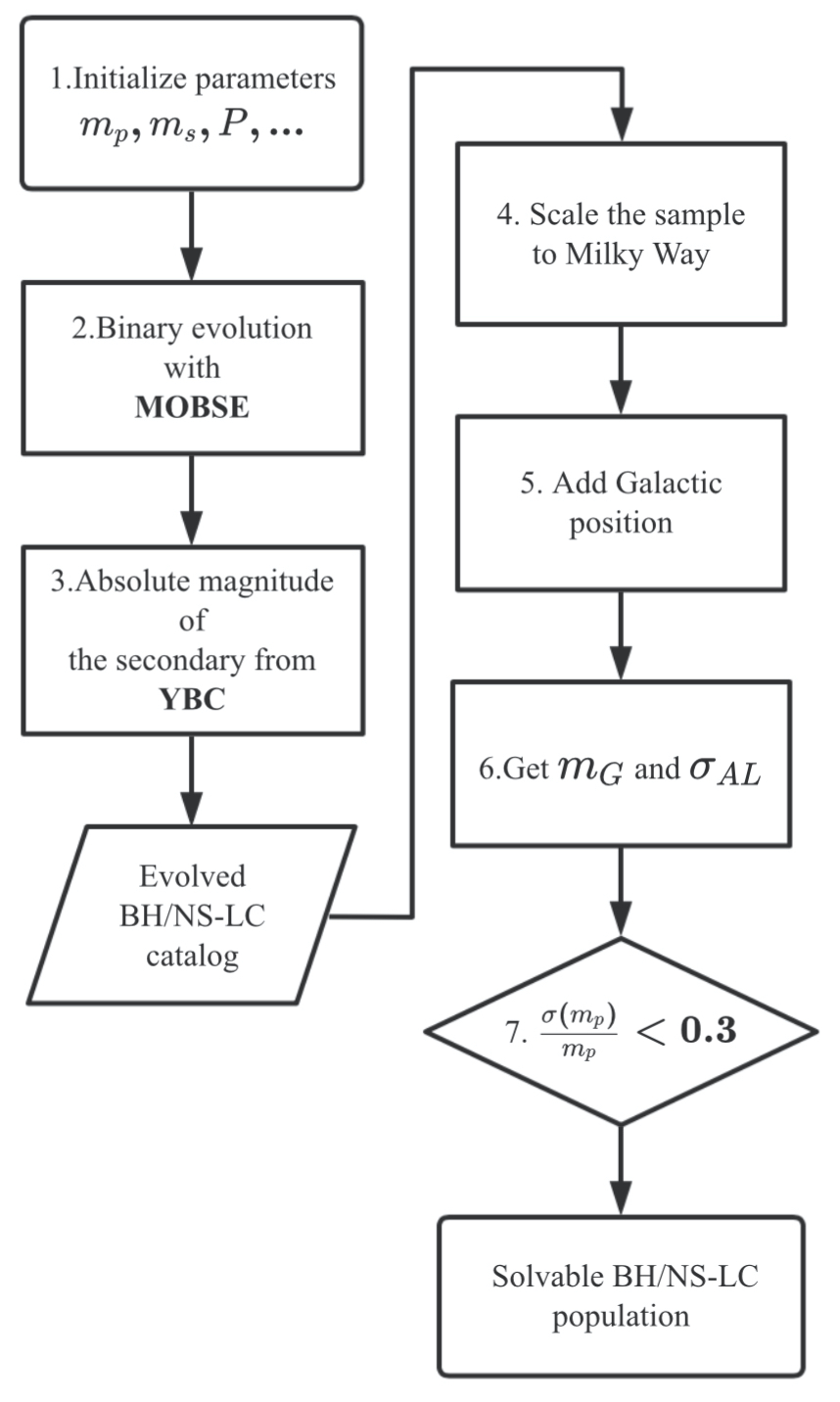}
\caption{Process of constructing the Gaia solvable BH/NS-LC population.}
\label{fig:process_BHNS}
\end{figure}
In order to examine the real ability of Gaia, we applied our $\Xi$-relation to a mock sample of a Galactic BH/NS-LC population using the binary evolution code \textsf{MOBSE}\footnote{For more details, we refer to https://mobse-webpage.netlify.app/about/}, which is based on the code Binary Star Evolution (\textsf{BSE}) \citep{BSE}, with some important upgrades to the models, including natal kicks, stellar winds, and supernovae  \citep[e.g.][]{stellar_wind,kick1,supernovae,kick2}. 

In  \textsf{MOBSE}  we adopted the following models. For the stellar winds we applied a metalicity-dependent model (see \citealt{Belczynski_wind_2010} and \citealt{Chen_wind_2015}). During the common envelope (CE) phase, the CE efficiency parameter $\alpha$,  which usually
has a great uncertainty, is set to 3 for simplicity \citep[e.g.][]{stellar_wind,kick1,shikauchi2021}. Another CE parameter, the binding energy factor $\lambda$, is calculated by the process in Appendix A of \cite{ce_lambda}. For the supernova (SN) model, we took the delayed SN engine model by \cite{Fryer2012}. Finally, we used the model in \cite{kick2} to calculate the BH and NS natal kicks.

The whole process described in this section is shown in Fig.~\ref{fig:process_BHNS}, which can be divided into two part. First, from Step 1 to Step 3  we generated the initial parameters of the binaries and construct the evolved sample of BH/NS-LCs (Sect.~\ref{section:binary_ini}). Second, from Step 4 to Step 7 we scaled and put this evolved sample to the Milky Way, and applied the $\Xi$-relation (Sect.~\ref{section:galactic_realization}). We repeated the second part for 500 times and give the statistical result of the solvable BH/NS-LC sample (Sect.~\ref{section:prediction_result}).

\subsubsection{Evolved sample of BH/NS-LC}\label{section:binary_ini}
In Step 1 and Step 2 we evolved the binaries with \textsf{MOBSE}, which needs the input parameters, $[m_p, m_s, P, e, Z, \rm{age}]$. The mass of the primary $m_p$ is sampled from the  power-law initial mass function (IMF) in \cite{IMF}. The mass of the secondary $m_s$, a less massive companion, is the product of $m_p$ and mass ratio $q$, which is generated from a uniform distribution $q\in[0.08/m_p,1]$. We referred to the study of \cite{sana2012} to initialise the massive binary systems. We adopted a power-law period distribution $\log P{\propto}(\log P)^{-0.55}$ and a power-law eccentricity distribution $e{\propto}e^{-0.42}$. The eccentricity falls in the range $[0,0.9]$, while $\log P$ is in the range $[0.15,5.5]$, a larger range from \cite{sana2014}.

For the metallicity $Z$ and the age we followed O20, who considered different Milky Way components, such as thin disk, thick disk, halo, and bulge. We ignored the globular clusters because their total mass only accounts for 0.005\%-0.01\% of the Milky Way stellar mass \citep[e.g.][]{globular_mass1,globular_mass2}. A short introduction of the distributions of $Z$ and age are described here, based on O20.
\begin{enumerate}[   ]
\item\textbf{Thin disk}. The age of binaries in the thin disk is a uniform distribution in the range of $[0,10]\unit{Gyr}$;   the metallicity decreases evenly from $1\,Z_{\odot}$ to $0.1\,Z_{\odot}$. The total mass of the thin disk is 90\% of the mass of the whole stellar disk, which is $5.17\pm1.11\times10^{10}\,M_{\odot}$ \citep{disk_mass}.
\item \textbf{Thick disk}. We followed O20, and considered  that the thick disk   formed uniformly between $9\unit{Gyr}$ and 11Gyr ago with a constant metallicity of $0.25\,Z_{\odot}$. The thick disk only accounts for 10\% of the stellar disk. 
\item \textbf{Bulge}. Referring to the figures in \cite{bulge_simulation}, we assumed that the bulge can be devided into two parts. One part, containing 48\% of the mass, formed from 
$12\unit{Gyr}$ to $10\unit{Gyr}$ ago with $Z$ that increased from $0.1\,Z_{\odot}$ to $1\,Z_{\odot}$. The other part formed   $10\unit{Gyr}$ ago, maintaining the metallicity at $1.5\,Z_{\odot}$.
\item \textbf{Halo}. Since most of the mass of the halo is dark matter, the stellar mass is only about $1.4\times10^{9}\,M_{\odot}$ (out to 100 kpc) \citep{halo_mass}. As O20 did, we also set two specific metallicity values for the halo, $Z=0.01\,Z_{\odot},0.02\,Z_{\odot}$. The binaries with $0.01\,Z_{\odot}$ formed between $12\unit{Gyr}$ and $11\unit{Gyr}$ ago, while the binaries with $0.01\,Z_{\odot}$ formed between $11\unit{Gyr}$ and $10\unit{Gyr}$ ago.
\end{enumerate}

To generate the evolved sample, which is used in Sect.~\ref{section:galactic_realization}, we generated 7.5 million to 17.5 million initial binaries for different Galactic components. We list the evolved results with the input initial systems in Table~\ref{tab:BH_NS_for_prediction}.

\begin{table}
\centering                          
\begin{tabular}{c c c c }        
\hline\hline                 
Galactic Component & BH-LC & NS-LC & Initial System \\ 
\hline                    
Thin Disk  & 65526 & 633221 & 17.5m \\
Thick Disk & 40908 & 289951 &  7.5m \\
Bulge      & 30458 & 502623 &  15m\\
Halo       & 69459 & 439264 &  10m\\
\hline                                   
\end{tabular}
\caption{Number of BH/NS-LCs in the evolved sample and number of the initial systems. \label{tab:BH_NS_for_prediction}}
\end{table}

After the binary evolution, in Step 3, we added the absolute magnitude $G$ to the luminous component in each system, where we chose the \textsf{YBC} database \citep{YBC} for the bolometric correction. The method for obtaining $G$ with the \textsf{YBC} website\footnote{http://stev.oapd.inaf.it/YBC/} is described in Appendix~\ref{append:absolute_magnitude}.

\subsubsection{Galactic realisation}\label{section:galactic_realization}
We used the following process in Step 4 to scale our evolved BH/NS-LCs to the whole Milky Way. According to O20, a sample of 2.5 million binaries with a primary from a $5-150\,M_{\odot}$ IMF accounts for a $3.1\times 10^8\,M_{\odot}$ stellar population from a    $0.08-150\,M_{\odot}$ IMF. Here the binary fraction is fixed at 50\%. For a Galactic component $\mathbf{c}$, which stands for [thin disk, thick disk, bulge,halo], we get $N_{sample,\mathbf{c}}$, the number of BH/NS-LCs in component $\mathbf{c}$, using the   linear scaling relation
\begin{eqnarray}\label{eq:scaling_relation}
N_{binary,\mathbf{c}}&{=}&2.5\times10^6\cdot\frac{M_{\mathbf{c}}}{3.1\times10^8M_{\odot}}\nonumber\\
N_{sample,\mathbf{c}}&{=}&N_{binary,\mathbf{c}}\cdot\frac{N_{bh/ns-lc, \mathbf{c}}}{N_{ini,\mathbf{c}}},
\end{eqnarray}
where $M_\mathbf{c}$ is the total mass of component $\mathbf{c}$, and $N_{binary,\mathbf{c}}$ is the number of binary in component $\mathbf{c}$, whose   primary follows the $5{-}150\,M_{\odot}$ IMF; $N_{ini,\mathbf{c}}$ (Col.~4) and $N_{bh/ns-lc,\mathbf{c}}$ (Col.~2, Col.~3) are listed in Table~\ref{tab:BH_NS_for_prediction}. Here we used the same IMF to assign the mass for different Galactic components and do the sampling with replacements  since we only evolved part of the potential binary systems in each component. We repeated the Galactic realisation 500 times to get a better estimation  of the number of  solvable sources and evaluate the uncertainty of the sampling. 

In Step 5, we generated the position for each system using the following model. For the thin disk, thick disk, and bulge, we used a function $MC\_samp$ in the Compact
Object Synthesis and Monte Carlo Investigation Code (\textsf{COSMIC}) programme package  \citep{COSMIC}. For the $MC\_samp$ we referred to \cite{McMillan_milkyway}, who gives an axisymmetric model for the disks and bulge. For the halo stars, we used the dual stellar halo model in \citet{dual_halo} with a break radius at $17.2\unit{kpc}$. 

Using the absolute magnitude from Step 3 and the Galactic position from Step 5, we calculated the apparent magnitude and get the $\sigma_{AL}$ with Sect.~\ref{section:apparent_magnitude} in Step 6. Then we applied our $\Xi$-relation and derived $\frac{\sigma(m_p)}{m_p}$. All the sources that satisfy our criteria in Step 7, $\frac{\sigma(m_p)}{m_p}{<}0.3$, would be recorded in one realisation. The total sample of the 500   realisations  makes up the sample of solvable BH/NS-LC population. 

\subsection{Galactic distribution of the solvable BH/NS-LCs}\label{section:result_galactic_distribution}
\begin{figure}[h!]
\centering
\includegraphics[scale=0.185]{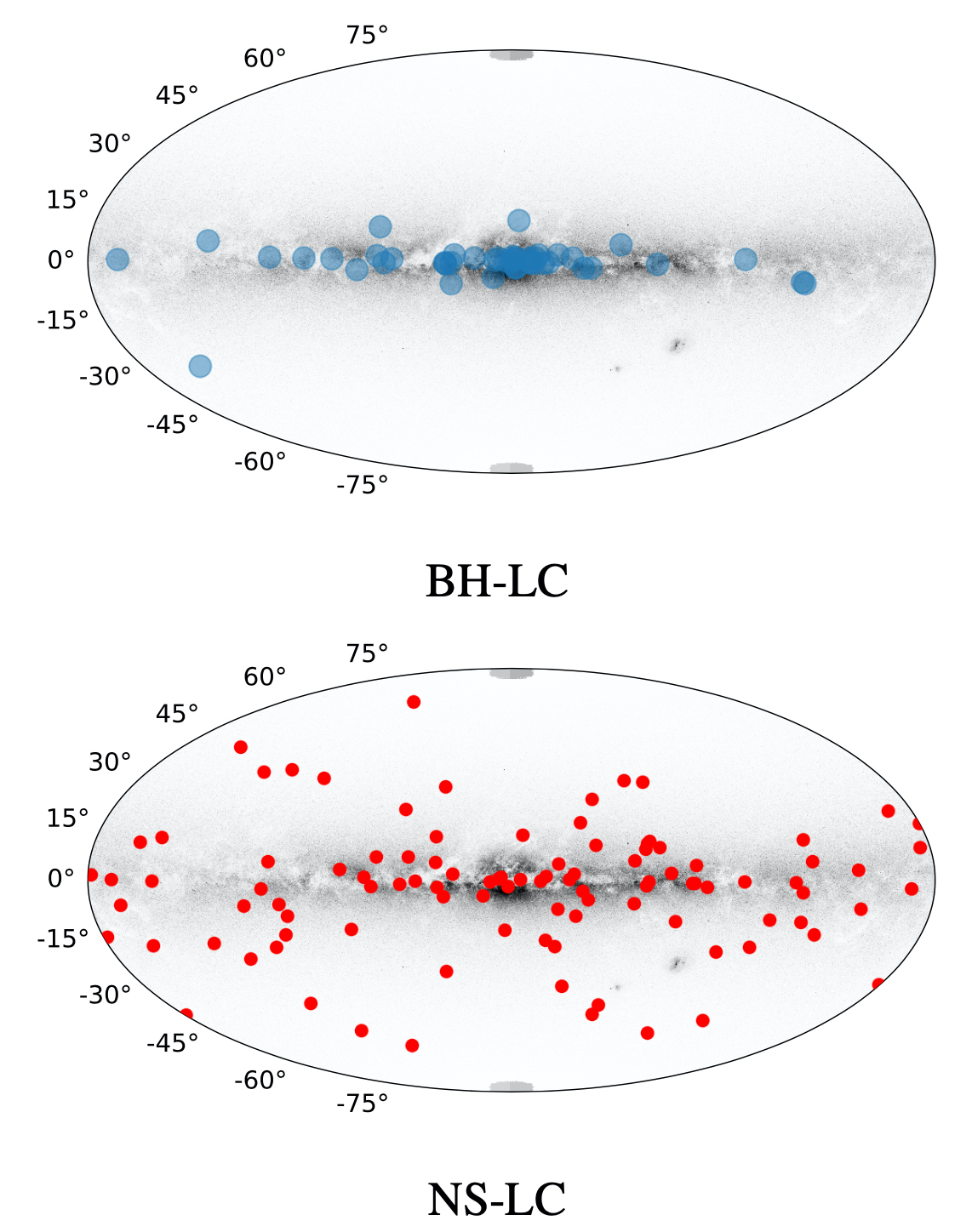}
\caption{One Galactic realisation of our simulated BH/NS-LC population with a real Milky Way background. The blue points in the \textbf{top} panel are the 48 BH-LCs, while the red points in the \textbf{bottom} panel are the  102 NS-LCs. These are the typical numbers for  Gaia solvable BH/NS-LCs in this work.
\label{fig:BH_NS_realization_sample}}
\end{figure}
We select a random realisation with 48 BH-LCs and 102 NS-LCs,  and plot them in Fig.~\ref{fig:BH_NS_realization_sample} with a real Milky Way background\footnote{We use the \textsf{mw\_plot} (https://pypi.org/project/mw-plot/) to plot our data and background, which is modified from an all-sky figure of ESA/Gaia/DPAC.}. It is clear that BH-LCs mainly fall on the disk\footnote{The result is different from the observation of BH  low-mass X-ray binaries (LXMBs), which is explained by the observation selection effect \citep{Jonker}.}, while the NS-LCs disperse much more to all of the sky.

\begin{figure*}[h!]
\includegraphics[scale=0.71]{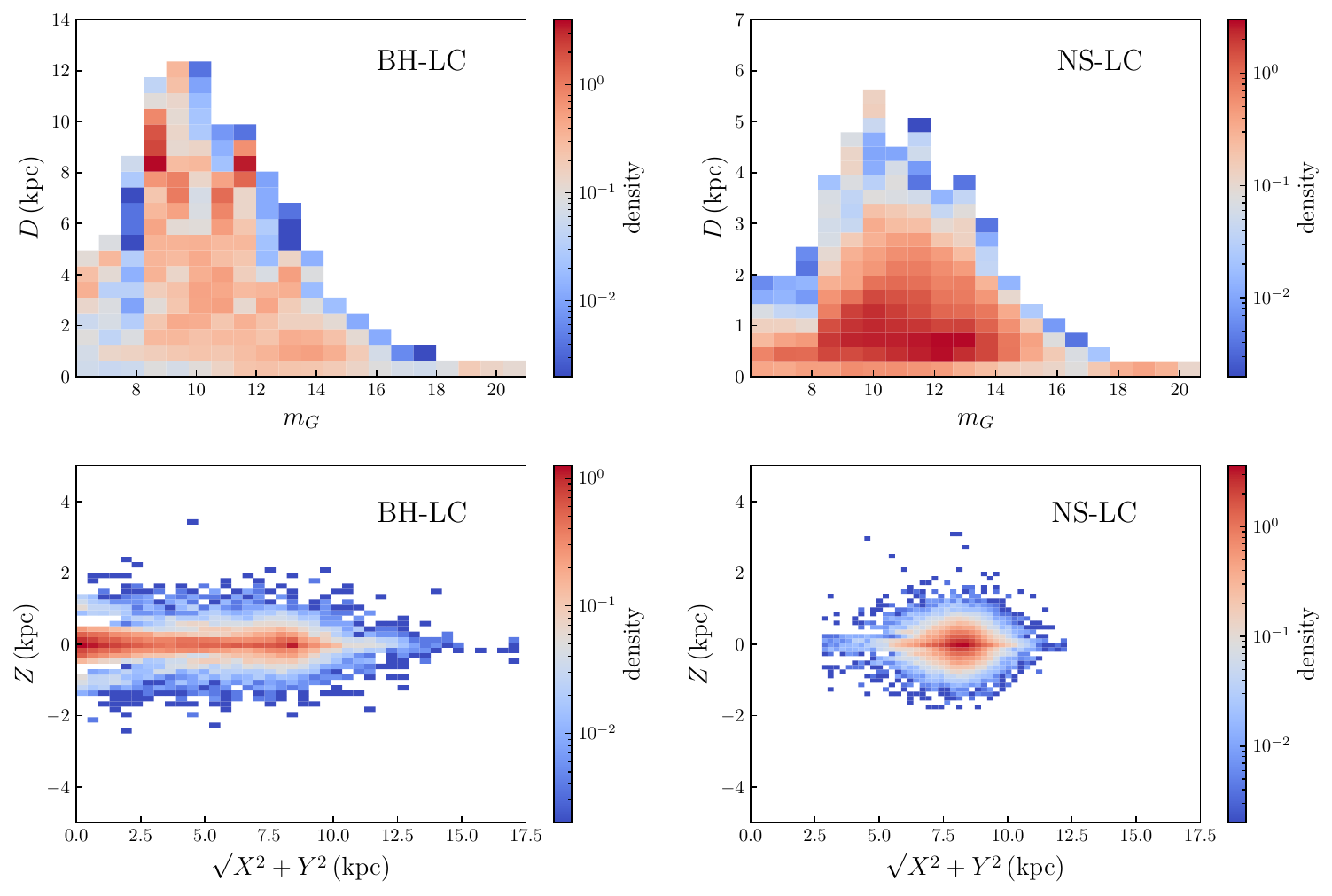}
\caption{Position and apparent magnitude distributions of BH-LCs and NS-LCs. \textbf{Top}: Two-dimensional distribution of the solvable BH/NS-LCs in the  $m_G{-}D$ plane. \textbf{Bottom}: Galactic distribution of the solvable BH/NS-LCs in the $\sqrt{X^2+Y^2}-Z$ plane. The colour stands for the density of the solvable systems appearing in each bin. \label{fig:gmag_dis_pos}}
\end{figure*}
We also provide the solvable domain for the realistic simulated Galactic BH/NS-LC population in Fig.~\ref{fig:gmag_dis_pos}. In the top panel most of solvable sources fall in the $m_G$ range of 8 to 14. In the bottom panel the Galactic solvable area is different for BH-LCs and NS-LCs. The NS-LCs and BH-LCs both mainly concentrate around the disk. The NS-LCs all fall in the area around the Solar System in $5\unit{kpc}$, mostly in $3\unit{kpc}$, while the BH-LCs distribute between the Solar System and the Galactic centre, clustering at the two ends. The BH-LCs could be observed  in the dense Galactic centre at a similar frequency to the neighbourhood of Solar System because the BH-LCs in the centre usually have a larger $a_s$, a brighter secondary, and a higher number density. 

\subsection{Discussion of the solvable BH/NS-LCs}\label{section:prediction_result}

\begin{table*}[ht!]
\renewcommand{\arraystretch}{1.5}
\centering          
\begin{tabular}{c c c c c c }     
\hline\hline
Binary Type & Observation time & $\frac{\sigma(m_p)}{m_p}$ & $N_{1}$ & $N_{2}$ & 
$N_{3}$ \\ 
\hline                    
BH-LC  & $5\unit{yr}$ & ${<}30\%$ & $48_{-7}^{+7}$ & $66_{-8}^{+7}$ & $224_{-16}^{+12}$ \\
BH-LC  & $5\unit{yr}$ & ${<}10\%$ & $2_{-1}^{+2}$ & $3_{-1}^{+2}$ & $8_{-2}^{+3}$ \\
\hline
NS-LC  & $5\unit{yr}$ & ${<}30\%$ & $102_{-10}^{+11}$ & $168_{-11}^{+13}$ & $556_{-22}^{+27}$ \\
NS-LC  & $5\unit{yr}$ & ${<}10\%$ & $7_{-3}^{+3}$ & $14_{-3}^{+5}$ & $37_{-5}^{+7}$\\
\hline 
BH-LC  & $10\unit{yr}$ & ${<}30\%$ & $108_{-10}^{+10}$ & $143_{-13}^{+12}$& - \\
BH-LC  & $10\unit{yr}$ & ${<}10\%$ & $4_{-2}^{+2}$ & $5_{-2}^{+3}$& - \\
\hline 
NS-LC  & $10\unit{yr}$ & ${<}30\%$ & $168_{-13}^{+15}$ & $281_{-17}^{+16}$& - \\
NS-LC  & $10\unit{yr}$ & ${<}10\%$ & $13_{-3}^{+4}$ & $25_{-5}^{+5}$& -\\
\hline                  
\end{tabular}
\caption{Predicted number of   Galactic solvable BH/NS-LCs with Gaia. The statistical result is discussed in Sect.~\ref{section:prediction_result}. Columns~1-3 list the basic information of the sample, from which we calculate $[N_1,N_2,N_3]$ with three different methods. $N_1$ is our result, while $N_2$ and $N_3$ are for comparison. All    three methods use Eq.~\ref{eq:mass_error_propagation} to obtain $\sigma(m_p)/m_p$, but  different $\Xi$ values. For $N_1$, the $\Xi$ is calculated from Eq.~\ref{eq:full_relation}, while we set the $\Phi_4(P,e,i)=1$ when we calculate the $\Xi$ for $N_2$. For $N_3$, we use Eq.~\ref{eq:and_5yr_KF} to get the $\Xi$, and thus the binary period and the observation time are limited within $5\unit{yr}$. \label{tab:Solvables_collection}}
\end{table*}

In this work we use $\frac{\sigma(m_p)}{m_p}{<}30\%$ as the criteria for solvable BH/NS-LCs, and ignore the possible obstruction in some dense areas. Table~\ref{tab:Solvables_collection} gives the number of BH/NS-LCs that are used in the  comparison in this section. The $N_1$ column lists the predicted number of BH/NS-LCs with $30\%$ or $10\%$ precision in a $5\unit{yr}$ mission or a $10\unit{yr}$ mission, calculated
by Eq.~\ref{eq:full_relation} and Eq.~\ref{eq:mass_error_propagation}. In a $5\unit{yr}$ Gaia mission, the number of solvable BH-LCs and NS-LCs are $48_{-7}^{+7}$ and $102_{-10}^{+11}$, respectively, of which 93\% BH-LCs and 97\% NS-LCs have periods $P{<}5\unit{yr}$. This result is consistent with the number range of 40-340 BH-LCs in \cite{greg1}. If the mission lasted for 10 yr, the number of solvable BH-LCs and NS-LCs would increase to 108 and 168. In both the $5\unit{yr}$ mission and the $10\unit{yr}$ mission, about 4\% systems of the solvable BH-LCs have a precision $\frac{\sigma(m_p)}{m_p}{<}0.1$, while 7\% of the solvable NS-LCs have this level of precision. 

As a comparison, we calculated $N_2$ in the same way as $N_1$, but set the $\Phi_4(P,e,i){=}1$ in Eq.~\ref{eq:full_relation}, which means $N_2$ is free from the influence of $[e,i]$. We find 66 BH-LCs and 168 NS-LCs in a $5\unit{yr}$ mission. Thus, our $\Phi_4(P,e,i)$ relation throws away 33\% of the BH-LCs and 38\% of the NS-LCs. This indicates that it is very important to do a complete simulation study of eccentricity and inclination before we can derive any statistical conclusion from the future observation result. For $N_3$ we used Eq.~\ref{eq:and_5yr_KF}, the result of A19, to get $\Xi$ and calculated $\frac{\sigma(m_p)}{m_p}$. It is also free from different $[e,i]$, so we compared it with $N_2$. In a $5\unit{yr}$ mission, $N_3$ predicts 2.3 times more   BH/NS-LCs ($\frac{\sigma(m_p)}{m_p}{<}30\%$), due to the difference between Eq.~\ref{eq:full_relation} and Eq.~\ref{eq:and_5yr_KF}. Here we ignored the sources with periods longer than $5\unit{yr}$ in $N_2$. 

\begin{table}[h!]
\renewcommand{\arraystretch}{1.5}
\centering                          
\begin{tabular}{c c c c c c }         
\hline\hline                 
Type & $\Xi$ & A19 & This work \\ 
\hline                    
BH-Giant  & ${<}30\%$ & $74\pm9$ & $16_{-4}^{+4}$ \\
BH-Giant  & ${<}10\%$ & $45\pm7$ & $1_{-1}^{+2}$ \\
\hline
NS-Giant  & ${<}30\%$ & $190\pm14$ & $25_{-5}^{+5}$ \\
NS-Giant  & ${<}10\%$ & $90\pm9$ & $4_{-2}^{+2}$\\
\hline                                   
\end{tabular}
\caption{Predicted solvable BH/NS-Giants from A19 and this work. We apply the Eq.~\ref{eq:and_5yr_KF} to our BH/NS-LC evolved sample and in Col.~4 we give the number  of solvable BH or NS with a giant as companion. The results of A19 are listed in Col.~3 for comparison.\label{tab:A19_comparison} }
\end{table}

In addition, we applied the same criteria as A19 did, $\Xi{<}30\%$, to the BH/NS-Giants in our evolved sample, and list the result in Table~\ref{tab:A19_comparison}. We got $16{\pm}4$ solvable BH-Giants and $25{\pm}5$ solvable NS-Giants in our evolved sample, while A19 predicted 74 solvable BH-Giants and 190 NS-Giants. In the solvable sources of this work, only about 1/16 BH-Giants and 4/25 NS-Giants can reach $\Xi{<}10\%$, while these fractions are much higher in A19, reaching 45/74 and 90/190, respectively.

We propose three reasons for this difference. The main reason might be that we used a different parameter distribution for the  binary initialisation. Our values  are different not only in the binary system parameters, such as period, eccentricity, mass ratio, and binary fraction, but also in   metallicity and age. The second reason is that we used a Milky Way model combining different observation and simulation analyses, which are different from the simulated Milky Way-like galaxy m12i  \citep{m12i} used in A19. The third reason might be our different  treatment of the bolometric correction and extinction, which have a great influence on the $\sigma_{AL}$. From the above comparison, there is a big difference in the number ratio of the precision at 30\% and 10\% between A19 and our comparison sample result. Thus, it seems that the third reason also plays an important role. 

\section{Conclusion}\label{section:conclusion}
We studied Gaia's ability to derive the mass of the dark component in BH/NS-LC binary systems, using a realistic MCMC astrometry simulation. We extended the   $\Xi$-relation from $5\unit{yr}$ to $12.5\unit{yr}$, which is 2.5 times  a $5\unit{yr}$ mission lifetime. A $10\unit{yr}$ mission would improve the precision of mass measurement and obtain more BH/NS-LCs. 

Assuming we obtained the mass $m_s$ of the secondary,  we connected the relative error of the primary $m_p$ and real observable variables by an intermediate parameter $KF$, the Keplerian factor, in Eq.~\ref{eq:kepler_factor} and Eq.~\ref{eq:mass_error_propagation}.
 
Using a MCMC simulation method, we gave the $\Xi$-relation, which can be used to predict $\Xi$, the relative error of $KF$, with direct observation variables $[a_s,\sigma_{AL},N,P,e,i,\beta]$. Assuming we obtained the mass of $m_s$ by additional spectroscopic data, we were able to  predict the $\frac{\sigma(m_p)}{m_p}$ with Eq.~\ref{eq:mass_error_propagation}. The example in Sect.~\ref{section:summary_full_relation} shows that our $\Xi$-relation is more adaptable and precise than Eq.~\ref{eq:and_5yr_KF}, the relation in A19. In addition, we especially explored the influence of the parameters $[P,e,i,\beta]$, which could be used as the correction function for further statistical study of BH/NS-LCs with Gaia. The result in Sect.~\ref{section:prediction_result} shows that the eccentricity $e$ and inclination $i$ would influence about 38\% of the solvable sources. 

A Galactic BH/NS-LC population is generated by the evolution code \textsf{MOBSE}. In this process, we used a series of results from different observations or simulations. By applying our $\Xi$-relation to this BH/NS-LC population, we predict that $48_{-7}^{+7}$ BH-LCs and $102_{-10}^{+11}$ NS-LCs could be solved with a $5\unit{yr}$ Gaia mission, consistent with previous results \citep[e.g.][]{greg1,Andrews2019}. A comparison between the result of A19 and this work is given in Sect.~\ref{section:prediction_result}, showing the influence of using different models of the $\Xi$-relation, binary population, Milky Way evolution and other details. In addition, we provide the distribution of these systems in the Milky Way as an instruction for future searching project.

\begin{acknowledgement}
The authors sincerely thank the anonymous referee for their useful comments. The authors are also grateful for the precious suggestion and help from Robert Soria. Y.W. is supported by National Science Foundation of China (NSFC) under grant numbers 11988101/11933004, National Key Research and Development Program of China (NKRDPC) under grant numbers 2019YFA0405504/2019YFA0405000
, and Strategic Priority Program of the Chinese Academy of Sciences under grant number XDB41000000. S.L. is supported by the Youth Innovation Promotion Association CAS, the grants from the Natural Science Foundation of Shanghai through grant 21ZR1474100, and National Natural Science Foundation of China (NSFC) through grants 12173069, and 11703065. We acknowledge the science research grants from the China Manned Space Project with NO.CMS-CSST-2021-A12 and NO.CMS-CSST-2021-B10. A.O. is supported by the Polish National Science Center (NCN) grant Maestro (2018/30/A/ST9/00050).  N.G. is supported by European Union’s H2020 ERC Starting grant No. 945155–GWmining, Leverhulme Trust grant No. RPG-2019-350, and Royal Society grant No. RGS-R2-202004.
\end{acknowledgement}

\begin{appendix}
\section{Absolute magnitude}\label{append:absolute_magnitude}
In Sect.~\ref{section:binary_ini} we used the YBC database \citep{YBC} to perform the bolometric correction and calculate the absolute magnitude of the luminous companion in our evolved BH/NS-LCs. To utilise the YBC database from the website, we sliced our evolved sample into different grids of $[Mass, L, T_{eff}]$ (mass, luminosity, and effective temperature of the secondary), which contains 112,000 grids in total. We selected ten random companion stars in each grid, or we selected all of the stars if the grid contains fewer  than ten stars. Some of the grids are   empty if no star falls into them. This subsample, containing 21789 sources, is from  the \textsf{YBC} website and we used the table for $M_G$. For each source $i$ in the total sample, we selected a source $j$ in the subsample that  falls in the same grid of $(Mass, L, T_{eff})$ and has the closest $T_{eff}$ to source $i$. We then   calculated the absolute magnitude $M_{G,i}$ as 
\begin{eqnarray}
M_{G,i}=-2.5\log_{10}(\frac{L_i}{L_j})+M_{G,j}
\label{eq:Absolute_magnitude}
,\end{eqnarray}
where the $M_{G,j}$ is obtained from the \textsf{YBC} website. 
\end{appendix}
%
%

\bibliographystyle{aa}
\bibliography{wyl2021a}

\end{document}